\documentclass[12pt,english]{article}

\usepackage[a4paper,lmargin=70pt,rmargin=70pt]{geometry}
\usepackage{setspace}
\usepackage{babel}
\usepackage{graphicx}
\usepackage{amsmath}
\usepackage{amssymb}
\usepackage{subcaption}
\usepackage{cite}
\usepackage[colorlinks=true,urlcolor=blue,anchorcolor=blue,citecolor=blue,filecolor=blue,linkcolor=blue,menucolor=blue,pagecolor=blue,linktocpage=true]{hyperref}

\newcommand{\dho}{\partial}
\newcommand{\Hence}{\quad\Longrightarrow\quad}

\newcommand{\ed}{\,.}
\newcommand{\ec}{\,,}
\newcommand{\ecq}{\ec\quad}
\newcommand{\dt}{\delta t}
\newcommand{\dX}{\delta X}
\newcommand{\dV}{\delta V}

\newcommand{\dx}{\delta x}

\newcommand{\phys}{\mathrm{phys}}
\newcommand{\pb}{\mathrm{p.b.}}

\newcommand{\cN}{\ensuremath{\mathcal{N}}}

\newcommand{\cM}{\ensuremath{\mathcal{M}}}

\DeclareMathOperator{\trace}{Tr}

\begin{document}

\begin{flushright} 
\today\\
YITP-15-97

SU-ITP-15/17

\end{flushright} 

\vspace{0.1cm}

\begin{center}
  {\LARGE
  
  Chaos in Classical D0-Brane Mechanics
}
\end{center}
\vspace{0.1cm}
\vspace{0.1cm}
\begin{center}
	
      Guy G{\sc ur-Ari}$^a$\footnote
      {E-mail: guyga@stanford.edu},  
      Masanori H{\sc anada}$^{abc}$\footnote
      {E-mail: hanada@yukawa.kyoto-u.ac.jp}   
and  
      Stephen  H. S{\sc henker}$^a$\footnote
      {E-mail: sshenker@stanford.edu}
          
\vspace{0.5cm}
	
$^a${\it Stanford Institute for Theoretical Physics,\\
Stanford University, Stanford, CA 94305, USA}

\vspace{0.2cm}

$^b${\it Yukawa Institute for Theoretical Physics, Kyoto University,\\
Kitashirakawa Oiwakecho, Sakyo-ku, Kyoto 606-8502, Japan}	

\vspace{0.2cm}

$^c${\it The Hakubi Center for Advanced Research, Kyoto University,\\
Yoshida Ushinomiyacho, Sakyo-ku, Kyoto 606-8501, Japan}

\end{center}

\onehalfspacing

\vspace{1.5cm}

\begin{center}
  {\bf Abstract}
\end{center}

We study chaos in the classical limit of the matrix quantum mechanical system describing D0-brane dynamics.
We  determine a precise value of the largest  Lyapunov exponent, and, with less precision, calculate the entire spectrum of Lyapunov exponents.  We verify that these approach a smooth limit as $N \rightarrow \infty$. We show that a classical analog of scrambling occurs with fast scrambling scaling, $t_* \sim \log S$.     These results confirm the $k$-locality property of matrix mechanics discussed by Sekino and Susskind. 

\vspace{0.1cm}

% Don't indent paragraphs
\setlength{\parindent}{0pt}
\setlength{\parskip}{1ex}

\newpage

\tableofcontents
%%%%%%%%%%%%%%%%%%%%%%%%%%%%%%%%%%%%%
%%%%%%%%%%%%%%%%%%%%%%%%%%%%%%%%%%%%%
%%%%%%%%%%%%%%%%%%%%%%%%%%%%%%%%%%%%%
\section{Introduction and Summary of Results}
%%%%%%%%%%%%%%%%%%%%%%%%%%%%%%%%%%%%%
%%%%%%%%%%%%%%%%%%%%%%%%%%%%%%%%%%%%%
%%%%%%%%%%%%%%%%%%%%%%%%%%%%%%%%%%%%%

This paper  is devoted to a study of  classical chaos in the classical limit of the matrix quantum mechanical system describing D0-brane dynamics.   In particular we compute Lyapunov exponents in this system. 

 The motivation for this work flows from 
 recent progress on the overlap between quantum chaos and quantum gravity.    These developments have their origin  in Quantum Information theory, and specifically in work done making good approximations to random unitary operators
\cite{dankert:2009a,harrow:2009a,arnaud:2008a,brown:2010a,diniz:2011a,Brown:2012gy}.  Such approximations can be implemented very quickly,  in a time proportional to $\log n$, where $n$ is the number of qubits (the analog of the entropy $S$ in this system).

Hayden and Preskill \cite{Hayden:2007cs} connected this timescale to one characteristic of black hole horizons \cite{'tHooft:1990fr,Kiem:1995iy} $t_* \sim R \log (M/m_p) \sim R \log S$, where $R$ is the Schwarzschild radius, $M$ is the black hole mass, $m_p$ is the Planck mass and $S$ is the Bekenstein-Hawking entropy of the black hole.  This logarithm is a consequence of the exponential blueshift of modes at late Schwarzschild time near the horizon, following from its Rindler structure.   They presented an example of a model typical of those  discussed in Quantum Information: a Hamiltonian coupling pairs of qubits nonlocally with a random pattern, with the 2-qubit gates being chosen at random.  It is easy to see that such a Hamiltonian will cause all qubits to become entangled with each other in a time of order $\log n$, and reasonable to conjecture that chaos has set in by this time \cite{dankert:2009a,harrow:2009a,arnaud:2008a,brown:2010a,diniz:2011a,Brown:2012gy,Hayden:2007cs}.  This conjecture is supported by analysis of quantum circuits and a Lieb-Robinson bound \cite{Lashkari:2011yi}.  A crucial aspect of such Hamiltonians is ``$k$-locality,'' where interactions can be nonlocal but only a finite number $k$ of qubits are coupled together in each term of the Hamiltonian, independent of the total number of qubits in the system.

Sekino and Susskind made the connection between these ideas and gauge/gravity duality \cite{Sekino:2008he}.  They argued that matrix quantum systems behave similarly to $k$-local qubit systems: the matrix indices function like the qubit label, and the sum over indices in the matrix interactions couples a finite number of index pairs together nonlocally, but satisfying the $k$-local property.  In some ways the simplest such system is maximally supersymmetric matrix quantum mechanics \cite{deWit:1988ig}, which has M-theory as its infrared gravity dual in Matrix Theory \cite{Banks:1996vh} and type IIA string theory at somewhat higher energies \cite{Itzhaki:1998dd}. The horizons of the black hole duals in such systems are Rindler in nature, and so matrix quantum systems have the characteristic logarithmic time which they interpreted as a ``scrambling time"  $t_* \sim \beta \log S$ (here $\beta \sim R$ is the inverse Hawking temperature of the black hole).
 Sekino and Susskind went on to make the ``fast scrambling conjecture,'' that in all reasonable physical systems chaos cannot set in faster than the logarithmic rate it does in black holes, in a time $t_* \sim \beta \log S$.   
 
 The next stage in the analysis of this kind of quantum chaos was undertaken in \cite{Shenker:2013pqa,Shenker:2013yza,Shenker:2014cwa,Roberts:2014isa,Kitaev_talk_2014,Maldacena:2015waa,Kitaev_talk_2015} using holographic (and other)  techniques.    
 A sharp diagnostic of chaos is the growth of a commutator \cite{Larkin1969,Almheiri:2013hfa} of simple operators with time, $C(t) = - \langle [V, W(t)]^2 \rangle$, where the brackets denote thermal expectation value.  In a chaotic quantum system $W(t)$ (in these contexts sometimes referred to as a ``precursor" \cite{Polchinski:1999yd})  becomes more complicated with time due to the lack of cancellation between the first and last factors in $W(t) = e^{iHt} W e^{-iHt}$ induced by the small perturbation $W$.   On expanding out the commutator one finds that the quantity most sensitive to chaos is an out-of-time-order correlator,
 $D(t) = \langle V W(t) V W(t) \rangle$.

As Larkin and Ovchinnikov \cite{Larkin1969} pointed out long ago,  in few body quantum systems described schematically by a coordinate $q$ and momentum $p$ the commutator $C(t) = \langle -[p, q(t)]^2 \rangle$ goes over in the semiclassical limit to 
$C(t) \rightarrow \hbar^2 \langle \{p, q(t)\}^2 \rangle$ where $\{\cdot,\cdot\}$ is the Poisson bracket.   This can be expressed as $\hbar^2 \Big\langle\left( \frac{\partial q(t)}{\partial q(0)} \right)^2 \Big\rangle = \hbar^2 \langle e^{2\lambda_L t} \rangle $, where $\lambda_L$ is the Lyapunov exponent. This motivates using the commutator as a diagnostic of chaos.
 
  The quantities $C(t)$, $D(t)$  (and closely related thermofield double two-sided correlators) have been computed holographically in \cite{Shenker:2013pqa,Shenker:2013yza,Shenker:2014cwa,Roberts:2014isa,Kitaev_talk_2014}.   The essential bulk phenomenon is a high energy collision between the quanta created by $V$ and $W(t)$ near the horizon.   The perturbative strength of gravitational scattering in such a collision is of order $G_N \, s$ (in AdS units) where $G_N$ is Newton's constant and $s$ is the center of mass energy squared.   The center of mass energy is (up to order one constants)  $s = \frac{1}{\beta^2} \exp{\frac{2 \pi t}{\beta}}$ because of the Rindler nature of the horizon and the role of boundary time as Schwarzschild time.   
 In the Einstein gravity limit the first term surviving in the commutator is second order in $G_N \sim 1/N^2$, 
 \begin{equation} \label{commbeh}
 C(t) \sim \left( \frac{1}{N^2} \exp{\frac{2 \pi t}{\beta}} \right)^2 \ed
 \end{equation}
 This becomes of order one at 
 \begin{equation}\label{scrtime}
 t_* = \frac{\beta}{2 \pi} \log N^2 =  \frac{\beta}{2 \pi} \log S .
 \end{equation}
 This is the precise large $N$ holographic scrambling time for systems with a bulk Einstein gravity dual. 
 Kitaev \cite{Kitaev_talk_2014}, building on \cite{Larkin1969},  connected the exponential time behavior in \eqref{commbeh} to Lyapunov behavior in chaotic systems.   Here the Lyapunov exponent is given by $\lambda_L = \frac{2 \pi}{\beta} = 2 \pi T$.\footnote{
 Notice that the exponential growth in \eqref{commbeh} gives twice the Lyapunov exponent, because the commutator is squared.
 } This exponential behavior and the small $\frac{1}{N^2}$ prefactor are the ingredients determining the fast scrambling conjecture timescale.
 
 The authors of \cite{Maldacena:2015waa} were able to establish the Einstein gravity value $\lambda_L = \frac{2 \pi}{\beta}= 2 \pi T$ as a sharp upper bound on thermal quantum systems with a large number of degrees of freedom and a large hierarchy between scrambling and dissipation times. The argument uses only general principles of quantum mechanics and plausible physical assumptions about the decay of time-ordered correlators.
 
 This bound does not enable one to compute the value of Lyapunov exponents in a given quantum system. 
  A suggestion was made in \cite{Shenker:2014cwa} about how to compute $\lambda_L$ at weak coupling, motivated by the BFKL ladder summation for  high energy scattering.  
  Stanford \cite{Stanford_talk_2015} has recently succeeded in implementing  this calculation in matrix $\phi^4$ quantum field theory.

 Kitaev \cite{Kitaev_talk_2015} has shown how to compute $\lambda_L$ in a strongly coupled large $N$ fermionic quantum mechanics system related to the Sachdev-Ye model \cite{Sachdev:1992fk,Sachdev:2015efa}.  He proceeds by summing ladder diagrams that in this case give the exact large $N$ solution to the model.  In the limit of strong coupling the exponent saturates the bound --- a remarkable result.

 Direct numerical work on this aspect of quantum gauge systems seems challenging.   Here we follow a different approach, exploring the {\it classical} dynamics of such a system.  In particular we explore the classical dynamics of the maximally supersymmetric matrix quantum mechanics in 0+1 dimensions, in the large $N$ limit.
The Lagrangian is
\begin{align}
  L =
  \frac{1}{2g^2}{\rm Tr}\left(
  \sum_{i}(D_tX^i)^2
  +
  \frac{1}{2}
  \sum_{i\neq j}[X^i,X^j]^2
  \right) + \cdots \ed
  \label{L}
\end{align}
Here $X^i$ $(i=1,\dots,9)$ are $N\times N$ traceless Hermitian matrices and
$D_tX^i = \partial_t X^i-[A_t,X^i]$ is the covariant derivative, where $A_t$ is the $SU(N)$ gauge field. 
We take the large $N$ limit with the `t Hooft coupling $\lambda = g^2 N$.
The remaining terms in \eqref{L} involve fermions, which do not contribute in the classical limit.

There have been a number of previous studies of the classical dynamics of this system, including \cite{Matinyan:1981dj,Savvidy:1982wx,Savvidy:1982jk,Asplund:2011qj,Asplund:2012tg,Asano:2015eha,Aoki:2015uha}.
Chaos was explored in \cite{Savvidy:1982wx,Savvidy:1982jk,Aref'eva:1997es,Aref'eva:1998mk,Asplund:2011qj,Asplund:2012tg,Asano:2015eha,Aoki:2015uha}.
In particular, \cite{Asplund:2011qj,Asplund:2012tg,Aoki:2015uha} studied the decay in time of two-point functions, 
and \cite{Aref'eva:1997es} studied the Lyapunov behavior.

At large $N$ and low temperature, the theory under discussion is holographically dual to a black hole in classical gravity.
We will focus on the large $N$, high temperature classical limit, where the dual black hole description is no longer valid.   The dimensionless parameter $\lambda_{\rm eff} = \lambda / T^3$ characterizing the large $N$ dynamics goes to zero in this limit.
(Previous numerical studies confirmed that there is no phase transition which separates the low and high temperature regions in this theory \cite{Anagnostopoulos:2007fw}. We therefore expect some qualitative features of the black hole, such as fast scrambling, to survive at high temperature.)

The high temperature limit of a quantum mechanical system is well approximated by its classical dynamics. 
This statement is only true for quantum mechanics, not quantum field theory ---
high-temperature field theory does not have a good classical limit because of the UV catastrophe.
Indeed, in high-temperature quantum field theory the occupation numbers of typical field modes are of order one, while classical equations of motion approximate quantum fields with large occupation numbers.\footnote{We thank Douglas Stanford for emphasizing this to us.}

Previous numerical studies \cite{Asplund:2011qj,Asplund:2012tg,Aoki:2015uha} showed that for generic initial conditions the classical system thermalizes into what can be thought of as a bound thermal state of $N$ D0-branes.   In this work we compute the Lyapunov exponents of this system by solving the equations of motion numerically.
For the leading exponent we give a precise result, while for the spectrum of subleading exponents we get a semi-quantitative estimate.
The classical system has a phase space with dimension that is of order $N^2$ and has the same number of Lyapunov exponents.  At large $N$ we find that they converge to a continuous spectrum with a finite maximum value.   That the chaotic dynamics has a smooth large $N$ limit provides support for the $k$-locality of matrix interactions, as discussed by Sekino and Susskind \cite{Sekino:2008he}.
In particular we find that that the largest Lyapunov exponent $\lambda_L$ approaches a finite value in the large $N$ limit, $\lambda_L \to 0.292
\,\lambda_{\rm eff}^{1/4}\,T$.   Note that this is parametrically smaller than the bound $\lambda_L \le 2 \pi T$ established in \cite{Maldacena:2015waa} in the classical limit $\lambda_{\rm eff} \rightarrow 0$. This determines the fast scrambling time, $t_* \sim \frac{1}{\lambda_L} \log N^2$, confirming that this model is a fast scrambler.

In classical systems the Lyapunov exponents are related to the Kolmogorov-Sinai (KS) entropy, which measures the rate of growth of coarse-grained entropy when the system is far away from equilibrium.
Pesin proved that the KS entropy is equal to the sum of positive Lyapunov exponents, and this result allows us to compute the KS entropy in the matrix theory.
Our result that the Lyapunov spectrum converges to a smooth density at large $N$ implies that the KS entropy is proportional to $N^2$.

The paper is organized as follows.
In Section~\ref{model} we present the matrix model and describe its classical limit.
In Section~\ref{lyapunov} we review the classical theory of Lyapunov exponents, and explain how it applies to the classical matrix model.
The main difficulty here is in dealing with the gauge symmetry of the model.
In Section~\ref{largest} we present numerical results for the Lyapunov exponent in the large $N$ limit, using various methods to compute the exponent.
Then, in Section~\ref{sec:LyapunovSpectrum} we present the computation of the Lyapunov spectrum in this system.
Section~\ref{discussion} includes a discussion of the results, and several appendices present some technical details of the computation.

%%%%%%%%%%%%%%%%%%%%%%%%%%%%%%%%%%%%%
%%%%%%%%%%%%%%%%%%%%%%%%%%%%%%%%%%%%%
%%%%%%%%%%%%%%%%%%%%%%%%%%%%%%%%%%%%%
\section{D0-Branes at High Temperature}\label{model}
%%%%%%%%%%%%%%%%%%%%%%%%%%%%%%%%%%%%%
%%%%%%%%%%%%%%%%%%%%%%%%%%%%%%%%%%%%%
%%%%%%%%%%%%%%%%%%%%%%%%%%%%%%%%%%%%%
The model we consider is the low-energy effective theory that lives on a stack of $N$ D0-branes \cite{Witten:1995im}.
It can be obtained by dimensionally reducing super Yang-Mills in 9+1 dimensions to zero space dimensions.
This is a supersymmetric quantum mechanics with a $U(N)$ gauge symmetry and an $SO(9)$ global R-symmetry.
Its degrees of freedom include nine $N \times N$ Hermitian matrices $X^i_{ab}$, $i=1,\dots,9$, $a,b=1,\dots,N$, as well as 16 fermions $\psi_{ab}$ in the spinor representation of $SO(9)$, and a gauge field $A_{ab}$.
The action is
\begin{align}
  S = \frac{1}{2 g^2} \int \! dt \, \trace \left\{
  (D_t X^i)^2 + \frac{1}{2} [X^i,X^j]^2
  + \bar{\psi} D_t \psi 
  + \bar{\psi} \gamma^i [X^i,\psi]
  \right\} \ed
  \label{action}
\end{align}
The covariant derivative is $D_t = \dho_t - [A_t,\ \cdot\ ]$, and summation over repeated $SO(9)$ indices is implied.
In this work we take the matrices $X^i$ to be traceless because the trace mode is decoupled.
When the matrices $X^i$ are diagonal, their $N$ eigenvalues correspond to the positions of the D0-branes in 9-dimensional flat space.
Off-diagonal elements correspond to open string degrees of freedom that are stretched between different branes.

Let us take the large $N$ limit, keeping the 't Hooft coupling $\lambda = g^2 N$ fixed.
The coupling $\lambda$ is dimensionful, and at finite temperature $T$ we can define the dimensionless coupling $\lambda_{\rm eff} = \lambda / T^3$ which controls the size of loop corrections.
We will take the limit of small $\lambda_{\rm eff}$, which is the weak coupling / high-temperature limit where classical dynamics provides a good approximation.
There, the fermions do not contribute to the dynamics of $X^i$, so we can discard them \cite{Kawahara:2007ib}.
We choose the gauge to be $A_t = 0$.
Integrating out the gauge field leads to the Gauss law constraint,
\begin{align}
  \sum_i [X^i, V^i] &= 0 \label{gauss} \ecq
  V^i \equiv \dot{X}^i \ec
\end{align}
which should be preserved due to gauge invariance.
Fixing $A_t=0$ does not completely fix the gauge; the residual gauge freedom corresponds to global (\textit{i.e.} time-independent) $SU(N)$ transformations.

We will work in an ensemble with fixed energy $E$, and where the conserved angular momentum is set to zero.\footnote{
The linear momentum $\trace(\dot{X}^i)$ vanishes trivially due to the traceless condition.}
Averages in this ensemble will agree with thermal averages in the thermodynamic limit $N \to \infty$; the corresponding temperature $T$ is given as follows. 
The equipartition theorem for this system relates temperature, energy and number of degrees of freedom as 
\begin{align}
  \langle K \rangle = 2 \langle U \rangle = \frac{n_{\rm dof}}{2} T \ed
\end{align}
The total energy is $E=K+U$ where $K$ is the kinetic energy and $U$ is the potential energy.
$n_{\rm dof}$ is the number of physical degrees of freedom.
Naively the total number of degrees of freedom is $d(N^2-1)$, where $d=9$ is the number of matrices, 
but in accounting for the Gauss law constraint \eqref{gauss} and the residual gauge symmetry we have to subtract $(N^2-1)$. 
Furthermore, the conservation of the angular momentum ${\rm Tr}(X_i\dot{X}_j-X_j\dot{X}_i)$ should be taken into account, reducing the number of degrees of freedom by $d(d-1)/2$.
Therefore, 
\begin{align}
  E= \frac{3}{4}n_{\rm dof}T \ecq
  n_{\rm dof}= (d-1)(N^2-1)-\frac{d(d-1)}{2} = 8(N^2-1)-36\ed
  \label{equipartition}
\end{align}

In the weak coupling limit we can use the classical approximation to describe the real-time evolution of the system, at least for typical states at a given temperature.\footnote{
The classical approximation is valid when the energy quanta, which correspond to the open string masses in this case, are much smaller than the temperature. 
When branes and open strings form a typical thermal state, the typical open string mass (or equivalently the distance between branes) scales as $(\lambda T)^{1/4}$ \cite{Kawahara:2007ib}, and therefore the classical approximation is valid at weak coupling. 
When one considers special configurations like a sparse gas of D0-branes, the classical approximation is not valid. 
} 
Thermodynamic properties can then be computed using ergodicity, which we assume.
(Numerical results are consistent with this assumption.)
The scalar equation of motion in our gauge is
\begin{align}
  \ddot{X}^i &= \sum_j [X^j, [X^i,X^j]] \label{Xeom} \ed
\end{align}
Equations \eqref{gauss} and \eqref{Xeom} fully describe the system in the classical approximation.
Notice that the equations do not depend on the coupling.
Therefore, due to the form of the action \eqref{action}, classical observables may depend on the temperature and the coupling only through the combination $\lambda T = \lambda_{\rm eff} T^4$; the power of this combination is then determined by dimensional analysis.
From now on we set $T=1$ without loss of generality.

%%%%%%%%%%%%%%%%%%%%%%%%%%%%%%%%%%%%%
%%%%%%%%%%%%%%%%%%%%%%%%%%%%%%%%%%%%%
%%%%%%%%%%%%%%%%%%%%%%%%%%%%%%%%%%%%%
\subsection{Discretization} 
%%%%%%%%%%%%%%%%%%%%%%%%%%%%%%%%%%%%%
%%%%%%%%%%%%%%%%%%%%%%%%%%%%%%%%%%%%%
%%%%%%%%%%%%%%%%%%%%%%%%%%%%%%%%%%%%%

In order to study the time evolution numerically we discretize the equation of motion \eqref{Xeom} while preserving the constraint \eqref{gauss} exactly.
For this purpose we write the equations of motion as
\begin{align}
  \dot{X}^i(t) &= V^i(t) \ec \\
  \dot{V}^i(t) &= F^i(t) \equiv \sum_j [X^j(t), [X^i(t),X^j(t)]] \ed
  \label{eom2}
\end{align}
The discretized evolution with time step $\dt$ is taken to be of order $\dt^2$ \cite{Aoki:2015uha}.
It is given by
\begin{align}
  X^i(t + \dt) &= X^i(t) + V^i(t) \cdot \dt
  + F^i(t) \cdot \frac{\dt^2}{2} \ec \cr
  V^i(t + \dt) &= V^i(t) + \left( 
  F^i(t) + F^i(t + \dt)
  \right) \cdot \frac{\dt}{2} \label{disc} \ed
\end{align}
It is easy to check that this prescription preserves the Gauss law constraint, namely that if the constraint $\sum_i [X^i,V^i] = 0$ holds at time $t$, then under the evolution \eqref{disc} it also holds at $t+\dt$.\footnote{This can be seen by using the relation $\sum_i [X^i(t),F^i(t)] = 0$, which follows from the Jacobi identity.}
All that is left is to ensure that the initial conditions obey the constraint and have zero angular momentum.
We do this by initially setting $V^i = 0$ while taking $X^i$ to have random (Gaussian) matrix elements.

In order to control the discretization error after evolving for time $t$, we use two different time steps: $\dt = 10^{-4}$ and $\dt = 5 \cdot 10^{-4}$, and compare the results. We compared several quantities such as the norm of the perturbation 
$|\delta X|$, whose definition will be given later in this paper, as well as $\trace (X_i^2)$ and $\trace ([X_i,X_j]^2)$.
We found agreement for $t \lesssim 60$. A similar comparison with the same discretization has been performed 
previously; see Fig.~2 of \cite{Aoki:2015uha}.

%%%%%%%%%%%%%%%%%%%%%%%%%%%%%%%%%%%%%
%%%%%%%%%%%%%%%%%%%%%%%%%%%%%%%%%%%%%
%%%%%%%%%%%%%%%%%%%%%%%%%%%%%%%%%%%%%
\section{Lyapunov Exponents}
\label{lyapunov}
%%%%%%%%%%%%%%%%%%%%%%%%%%%%%%%%%%%%%
%%%%%%%%%%%%%%%%%%%%%%%%%%%%%%%%%%%%%
%%%%%%%%%%%%%%%%%%%%%%%%%%%%%%%%%%%%%
In this section we briefly review the theory of Lyapunov exponents in classical systems, and its application to the matrix model.
We stress the complexities that arise due to the gauge symmetry of our model.

Consider a Hamiltonian system with a phase space $\cM$ of dimension $n$.
Hamilton's equations define the mapping of a point $x_0$ in phase space to a point $x(t)$ after time $t$.
By linearizing Hamilton's equations we can define a linear operator $U(t;x_0)$ (the transfer matrix), that maps a tangent vector $\delta x_0$ (\emph{i.e.} an infinitesimal perturbation) at $x_0$ to a final vector $\delta x(t)$ at $x(t)$.

The signature of a chaotic system is the exponential growth of perturbations.
In order to discuss this growth we introduce a Riemannian metric $g$ on phase space.
In a chaotic system, a typical perturbation grows as $|\delta x(t)| \sim |\delta x_0| e^{\lambda_L  t}$, where $|\delta x| = \sqrt{g(\delta x,\delta x)}$. 
We define the Lyapunov exponent that is associated with the initial perturbation by
\begin{align}
  \lambda_L = \lim_{t \to \infty} \frac{1}{t}
  \log \left( \frac{|\delta x(t)|}{|\delta x_0|} \right) \ed
  \label{l}
\end{align}
Note that there is no natural choice for $g$ on phase space, but if phase space is compact then the Lyapunov exponents are independent of $g$; see Appendix \ref{indep}.
If phase space is noncompact then the exponents will not be well-defined in general.

In an ergodic system, the Lyapunov exponents $\lambda_L$ can take up to $\dim(\cM)=n$ distinct values \cite{oseledets1968multiplicative}.
The largest exponent is the one that is referred to as `the' Lyapunov exponent, because it dominates the growth of typical (non-fine-tuned) perturbations.
The spectrum of Lyapunov exponents is determined by the size of $g\left( U(t;x_0) \dx,U(t;x_0) \dx \right) = g\left( \dx,U^\dagger(t;x_0)U(t;x_0)\dx \right)$, namely by the eigenvalues of $U^\dagger(t;x_0)U(t;x_0)$.
Equivalently, the spectrum can be determined by performing a singular-value decomposition (SVD) on the transfer matrix;
here we choose an orthonormal basis for the tangent space (with respect to $g$), and write the transfer matrix in this basis as 
\begin{align}
  U(t;x_0) = W(t;x_0) \Sigma(t;x_0) V(t;x_0)^\dagger \ec \label{SVD}
\end{align}
where $W,V$ are unitary and $\Sigma = \rm{diag}(\sigma_1,\dots,\sigma_n)$ is positive-definite, with $\sigma_1 \ge \cdots \ge \sigma_n \ge 0$.
The Lyapunov exponents $\lambda_1,\dots,\lambda_n$ are then given in terms of the decomposition by
\begin{align}
  \lambda_i(x_0) = \lim_{t \to \infty} \frac{1}{t}
  \log \sigma_i(t;x_0) \ed
  \label{lambdai}
\end{align}
For ergodic systems, $\lambda_i = \lambda_i(x_0)$ is independent of the starting point $x_0$.
Phase space carries with it a symplectic structure (a closed, non-degenerate 2-form $\omega$), and the transfer matrix is a symplectic transformation.
Therefore, the Lyapunov exponents are paired: For every exponent $\lambda_i$ there is a corresponding exponent $-\lambda_i$ \cite{Xu20031}.
We will be interested in taking the limit in which the dimension of phase space $n$ goes to infinity (this will correspond to a `t Hooft limit of our matrix model).
As we will see, in the matrix model the set of discrete exponents $\lambda_i$ approach a distribution $\rho(\lambda)$ in this limit. The distribution is supported on $[-\lambda_L,\lambda_L]$ where $\lambda_L$ is finite. 

%%%%%%%%%%%%%%%%%%%%%%%%%%%%%%%%%%%%%
%%%%%%%%%%%%%%%%%%%%%%%%%%%%%%%%%%%%%
%%%%%%%%%%%%%%%%%%%%%%%%%%%%%%%%%%%%%
\subsection{Finite Time Approximation}
\label{sec:syserr}
%%%%%%%%%%%%%%%%%%%%%%%%%%%%%%%%%%%%%
%%%%%%%%%%%%%%%%%%%%%%%%%%%%%%%%%%%%%
%%%%%%%%%%%%%%%%%%%%%%%%%%%%%%%%%%%%%

In a numerical calculation of the exponent based on \eqref{l}, time must be kept finite.
We define the time-dependent exponents $\lambda_i(t;x_0)$ by
\begin{align}
  \lambda_i(t;x_0) \equiv \frac{1}{t} \log \sigma_i(t;x_0) \ecq
  i=1,\dots,n \ed
  \label{tlambda}
\end{align}
They converge to the Lyapunov exponents $\lambda_i$ as $t \to \infty$.
Due to the symplectic structure, the exponents are paired: $\lambda_i(t; x_0)$ and $-\lambda_i(t; x_0)$ appear together.

Let $\dx_0$ be a generic perturbation of unit norm.
Given the decomposition \eqref{SVD}, let $\{v_i(t)\}$ be the column vectors of $V(t)$ such that
\begin{align}
  U(t) v_i(t) = \sigma_i(t) w_i(t) \ec
\end{align}
where $w_i(t)$ are the columns of $W(t)$ (from now on the dependence on $x_0$ will be implicit).
Expand the initial perturbation as $\dx_0 = \sum_i c_i(t) v_i(t)$.
The evolved perturbation then has squared norm
\begin{align}
  \left| \dx(t) \right|^2 &= 
  \left| U(t) \dx_0 \right|^2 = 
  \sum_{i=1}^n |c_i(t)|^2 \sigma_i^2(t) \simeq 
  \frac{1}{n} \sum_{i=1}^n e^{2\lambda_i(t) t}
  \ed
\end{align}
In the last step we used the fact that for a typical vector $\dx_0$ we expect 
that $|c_i(t)|^2 \approx 1/n$.
The Lyapunov exponent (defined in \eqref{l}) is then approximated at finite times by
\begin{align}
  \lambda_L(t) \equiv
  \frac{1}{2t} \log \left( 
  \frac{1}{n} \sum_i e^{2\lambda_i(t) t} \right) 
  \ed
\end{align}
In Hamiltonian systems, it was argued that individual exponents typically approach their asymptotic values as $\lambda_i(t) \sim \lambda_i + \frac{a_i}{t}$ after averaging over initial conditions \cite{goldhirsch1987stability}.\footnote{
A heuristic way to understand the $1/t$ term is to assume that exponential behavior does not set in immediately at $t=0$, but only after a short transient.
}
In the matrix model, it will turn out that the individual exponent evolution is well-approximated by $\lambda_i(t) \sim \lambda_i + \frac{a_i}{t} + \frac{b_i \log t}{t}$.
We will also find that the effective exponent $\lambda_L(t)$ approaches its asymptotic value $\lambda_L$ much faster than do the individual exponents.

%%%%%%%%%%%%%%%%%%%%%%%%%%%%%%%%%%%%%%%%
%%%%%%%%%%%%%%%%%%%%%%%%%%%%%%%%%%%%%%%%
%%%%%%%%%%%%%%%%%%%%%%%%%%%%%%%%%%%%%%%%
\subsection{Matrix Model Application}
\label{mme}
%%%%%%%%%%%%%%%%%%%%%%%%%%%%%%%%%%%%%%%%
%%%%%%%%%%%%%%%%%%%%%%%%%%%%%%%%%%%%%%%%
%%%%%%%%%%%%%%%%%%%%%%%%%%%%%%%%%%%%%%%%
Let us now consider the Lyapunov exponents in the context of the D0-brane system.
The phase space $\cM$ of this system (after gauge fixing to $A_t=0$) is a vector space with coordinates $(X^i,V^i)$ and a symplectic structure that is given by $\omega = \sum dX^i_{ab} \wedge dV^i_{ba}$.
As explained above, in order to have well-defined Lyapunov exponents the space should be compact.
Let us restrict ourselves to a subspace with fixed energy.
This subspace is still noncompact due to the existence of flat directions: A configuration of the form
\begin{align}
  X^i = \begin{pmatrix}
    y^i & 0 & 0 \\
    0 & \cdot & \cdot \\
    0 & \cdot & \cdot
  \end{pmatrix}
  \label{sep}
\end{align}
has energy that is independent of the brane position $y^i$.
However, as we show in Appendix~\ref{appendix:flat_direction}, 
simple estimates suggest that even for finite $N$ the fixed-energy phase space has finite volume, and this is confirmed by the equilibration of the system even for $N=2$.\footnote{This is in contrast to the supersymmetric quantum system, where supersymmetric cancellation leads to actual flat directions.
In that system, entropic arguments at finite temperature show that the flat directions are exponentially suppressed at large $N$.
}
Therefore, the Lyapunov exponents are effectively well-defined for this classical system.

The next problem we face is that of gauge redundancy.
Having fixed the gauge to $A_t=0$, all physical configurations must satisfy the Gauss law constraint \eqref{gauss}.
Let us restrict our space to the constraint surface\footnote{
In a slight abuse of notation, in what follows we will use the velocity $V$ as momentum coordinates on phase space.
}
\begin{align}
  \cM_0 \equiv \Big\{ (X,V) \;\,\Big|\;\, \sum_i [X^i,V^i] = 0 \Big\} \ed
\end{align}
Restricting to $\cM_0$ is not sufficient because $\cM_0$ is not a phase space (in general it does not admit a symplectic structure), and also because of residual gauge symmetries.
To see this, let us define a Riemannian metric $g$ on the phase space $\cM$ by
\begin{align}
  g(\delta x, \delta x') = 
  g(\dX,\dV;\dX',\dV') \equiv
  \trace ( \dX \dX') + \trace(\dV \dV') \ed
  \label{g}
\end{align}
Here, $\delta x = (\dX,\dV)$ denotes a vector in phase space.
This metric is invariant under the residual gauge transformations (with respect to the gauge $A_t=0$)
\begin{align}
  X^i \to \tilde{U} X^i \tilde{U}^\dagger \ecq
  V^i \to \tilde{U} V^i \tilde{U}^\dagger \ecq
  \tilde{U} \in SU(N) \ed
  \label{res}
\end{align}
However, the metric \eqref{g} leads to a non-zero geodesic distance between gauge-equivalent configurations, namely between two configurations that are related by the transformation \eqref{res}.
Therefore, using the phase space $\cM$ (or the constrained space $\cM_0$) with the metric \eqref{g} to define the Lyapunov exponents will lead to `spurious' exponents that correspond to pure gauge modes rather than to physical perturbations.

The solution to this problem is to define a physical phase space from which the pure gauge modes have been modded out.
This procedure is known as the symplectic reduction of $\cM$, and it is explained in detail in Appendix~\ref{lyapapp}.
The upshot it that the physical Lyapunov exponents are obtained from a modified transfer matrix given by
\begin{align}
  U_{\phys}(t;x_0) \equiv P(x(t)) \cdot U(t;x_0) \cdot P(x_0) \ec
  \label{Uphys1}
\end{align}
where $P(x)$ is a projector that projects out vectors that do not obey the Gauss law constraint, as well as vectors that correspond to pure gauge transformations.
The gauge-invariant exponents are obtained as before by a singular value decomposition of $U_{\phys}$.

The presence of residual gauge transformations does not affect the leading exponent, essentially because perturbations corresponding to gauge transformations do not grow with time.
In the following section we will compute the leading exponent, so we will be able to ignore this issue.
In Sec.~\ref{sec:LyapunovSpectrum} we will compute the full spectrum of exponents, and there the prescription \eqref{Uphys1} will be used.

%%%%%%%%%%%%%%%%%%%%%%%%%%%%%%%%%%%%%%%%
%%%%%%%%%%%%%%%%%%%%%%%%%%%%%%%%%%%%%%%%
%%%%%%%%%%%%%%%%%%%%%%%%%%%%%%%%%%%%%%%%
\section{Leading Exponent Computation}\label{sec:leading_exponent}
\label{largest}
%%%%%%%%%%%%%%%%%%%%%%%%%%%%%%%%%%%%%%%%
%%%%%%%%%%%%%%%%%%%%%%%%%%%%%%%%%%%%%%%%
%%%%%%%%%%%%%%%%%%%%%%%%%%%%%%%%%%%%%%%%
In this section we compute the leading Lyapunov exponent of the classical matrix model by following diverging trajectories in phase space.
Our main result is that the exponent converges at large $N$.
One important corollary is that the classical matrix model is a fast scrambler, namely that the classical analog of scrambling time (defined below) scales as $\log N^2$.
Finally, we compute the exponent using an alternative method by considering gauge-invariant correlation functions, and find good agreement.

The computation of the Lyapunov exponent consists of three steps.
\begin{enumerate}
  \item `Thermalize' the system by evolving it for a long enough time.
  \item Perturb the system.
  \item Evolve both the original and the perturbed configurations, measuring the exponential rate at which they diverge.
\end{enumerate}
Let us discuss each step in detail.

\begin{figure}
  \centering
  \includegraphics[width=0.7\textwidth]{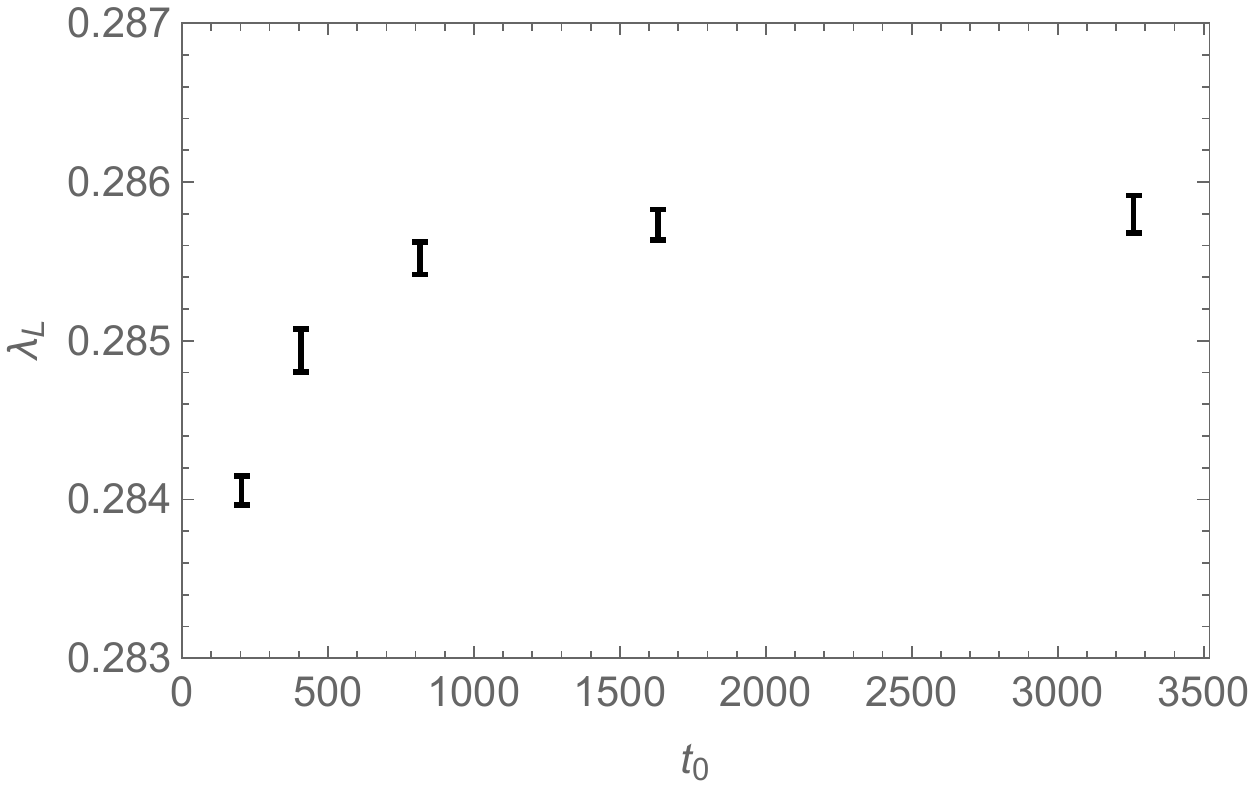}
  \caption{Lyapunov exponent as a function of thermalization time $t_0$.
  }
  \label{fig:lyap-therm}
\end{figure}

We begin by choosing an initial state where the $X$ variables are random and traceless, and where $\dot{X}=0$.
This initial state satisfies the Gauss law constraint, and also has vanishing momentum and angular momentum.
We then evolve the system for a sufficiently long time, so that it reaches a `typical state' that is uncorrelated with the (atypical) initial conditions.
This is the ergodic equivalent of thermalization.
How long do we need to evolve for in order to thermalize the system?
Fig.~\ref{fig:lyap-therm} shows the resulting Lyapunov exponents as a function of thermalization time $t_0$. 
(We will explain how the exponents are evaluated shortly.)
We see convergence for $t_0 \gtrsim 2000$, and in what follows we set $t_0 = 4000$.
Note that this is much longer than the thermalization time typically needed for other observables, and for observables previously studied in the literature; see e.g. \cite{Asplund:2011qj,Aoki:2015uha}.
The origin of this slow relaxation phenomenon is mysterious and is an interesting topic for future research.

Given a thermalized configuration $(X,V)$, we perturb it slightly while preserving the Gauss law constraint \eqref{gauss} 
by using the method described in Appendix~\ref{appendix:perturbation}. 
Having obtained the reference configuration $(X,V)$ and the perturbed configuration $(X',V')$, we evolve both together and compute the distance between them.
The distance function we use is
\begin{align}
  |\dX(t)| = \sqrt{ \sum_{i=1}^9 \trace( \dX_i^2(t) ) } \ecq
  \dX(t) \equiv X'(t) - X(t) \ed
  \label{dist}
\end{align}
The distance grows exponentially,
\begin{align}
  |\dX(t)| \sim |\dX(0)| e^{\lambda_L t} \ec
  \label{growth}
\end{align}
where $\lambda_L$ is the Lyapunov exponent.\footnote{
Note that we are using a naive distance function (similar to the one given by the metric \eqref{g}) which does not take into account gauge redundancy.
As explained in Sec.~\ref{mme}, this simplification does not affect the leading exponent.
The fact that we are considering only the distance only in $X$ and not in $V$ also does not change the resulting exponent.
}

\begin{figure}
  \centering
  \includegraphics[width=0.7\textwidth]{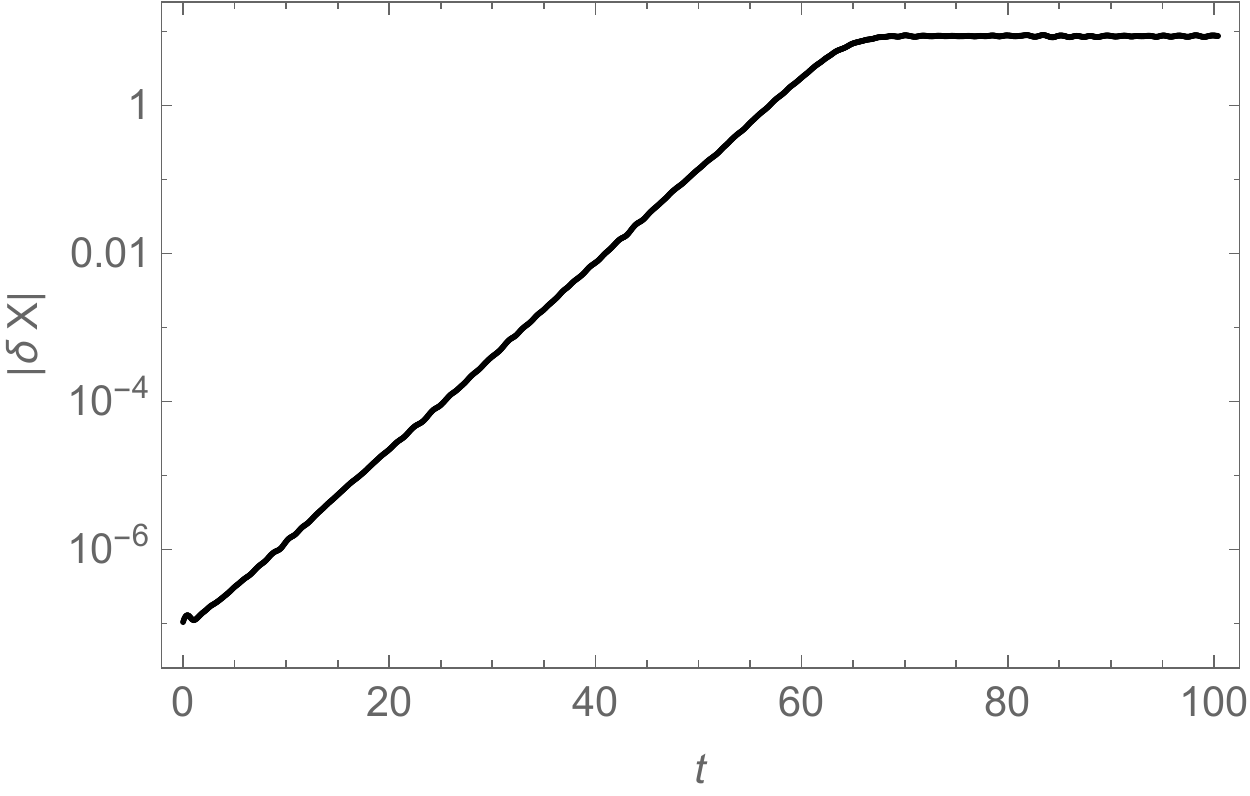}
  \caption{Time evolution of $|\dX(t)|$ for $N=16$. 
  Here $t=0$ is the time of the initial perturbation.}
  \label{fig:evol}
\end{figure}

\begin{figure}
  \centering
  \includegraphics[width=0.8\textwidth]{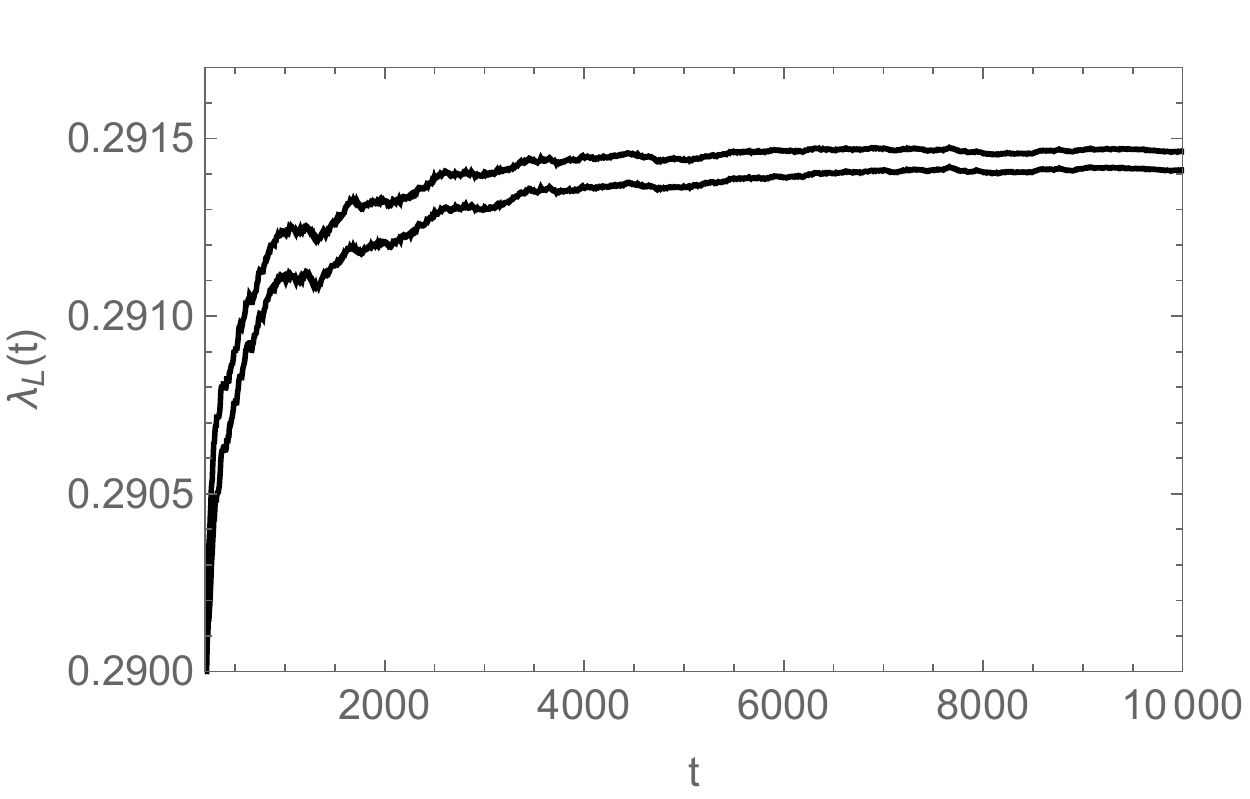}
  \caption{The exponent estimated using the Sprott algorithm as a function of time, for $N=20$ and at $\lambda_{\rm eff}^{1/4} \, T = 1$.
  The band represents the statistical fluctuations of different samples.}
  \label{fig:sprottvst}
\end{figure}

The evolution of $|\dX(t)|$ is shown in Fig.~\ref{fig:evol}.
Exponential growth sets in quickly, and continues until the size of the perturbation becomes of the order of the system size, at around $t \simeq 60$.
We shall call this the `scrambling time' $t_*$ of the perturbation.
In principle, the Lyapunov exponent can be extracted directly from the exponential growth.
As discussed in Sec.~\ref{sec:syserr}, the accuracy of this calculation is limited by the finite time of the perturbation growth.
For this reason we now consider Sprott's algorithm \cite{sprott2003chaos}, which is an alternative method for computing the exponent.
The algorithm is explained in Appendix~\ref{sprott}.
It allows us to study the growth at arbitrarily long time scale (we used $t=10^4$), and to extract the largest Lyapunov exponent.
Fig.~\ref{fig:sprottvst} shows the convergence of the exponent computed using this algorithm.
Notice that convergence to within one percent occurs at $t \lesssim 100$.
This suggests that the Sprott result should be close to the direct fit result, and this is indeed what we find.
In Sec.~\ref{sec:LyapunovSpectrum} we determine the subleading exponents and give more detail on how this agreement is achieved. 

\begin{figure}
  \centering
  \includegraphics[width=0.7\textwidth]{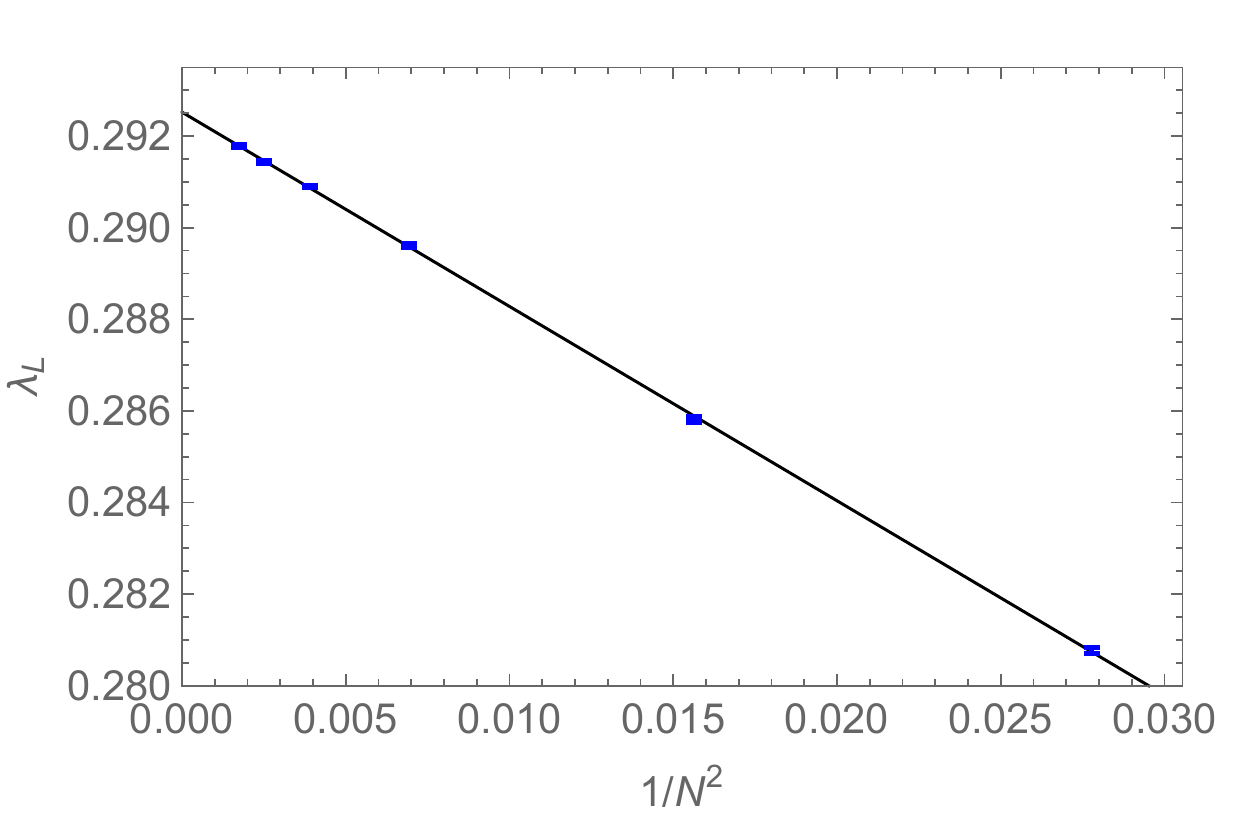}
  \caption{Leading Lyapunov exponent for $N=6,8,12,16,20,24$ at $\lambda_{\rm eff}^{1/4} \, T = 1$. The error bars are statistical.}
  \label{fig:lyapunov}
\end{figure}

The measured Lyapunov exponent for various values of $N$ is shown in Fig.~\ref{fig:lyapunov}.
We find that the large $N$ behavior is given by\footnote{
Note that the uncertainties quoted here do not take into account the error bars, so they are an underestimate.
}
\begin{align}
  \lambda_L = \left[ 0.29252(2)
  - \frac{0.424(2)}{N^2} \right]
  \lambda_{\rm eff}^{1/4} \, T \ed
  \label{lambda}
\end{align}
The dependence on the temperature and the coupling follows from dimensional analysis, as explained in Sec.~\ref{model}.
The fact that the leading correction goes as $N^{-2}$ is consistent with the 't Hooft counting. 

%%%%%%%%%%%%%%%%%%%%%%%%%%%%%%%%%%%%%%%%
%%%%%%%%%%%%%%%%%%%%%%%%%%%%%%%%%%%%%%%%
%%%%%%%%%%%%%%%%%%%%%%%%%%%%%%%%%%%%%%%%
\subsection{Fast Scrambling}
%%%%%%%%%%%%%%%%%%%%%%%%%%%%%%%%%%%%%%%%
%%%%%%%%%%%%%%%%%%%%%%%%%%%%%%%%%%%%%%%%
%%%%%%%%%%%%%%%%%%%%%%%%%%%%%%%%%%%%%%%%

Quantum systems have a natural notion of `scrambling time' $t_*$, which is the time it takes a local perturbation to completely de-localize, or become `scrambled'.
In classical systems we can only discuss the scrambling time of a given perturbation (rather than as a property of the system itself). 
This is because we can make the growth time of a perturbation arbitrarily large by taking the initial perturbation to be small (in quantum systems we are limited by uncertainty).
Earlier we defined the scrambling time to be the time at which $|\dX(t)|$ stops growing.
We can then consider the $N$ scaling of the scrambling time by scaling $N$ while keeping the size of the initial perturbation fixed (quantum mechanically, the minimal size of a perturbation is $O(N^0)$).

\begin{figure}
  \centering
  \includegraphics[width=0.7\textwidth]{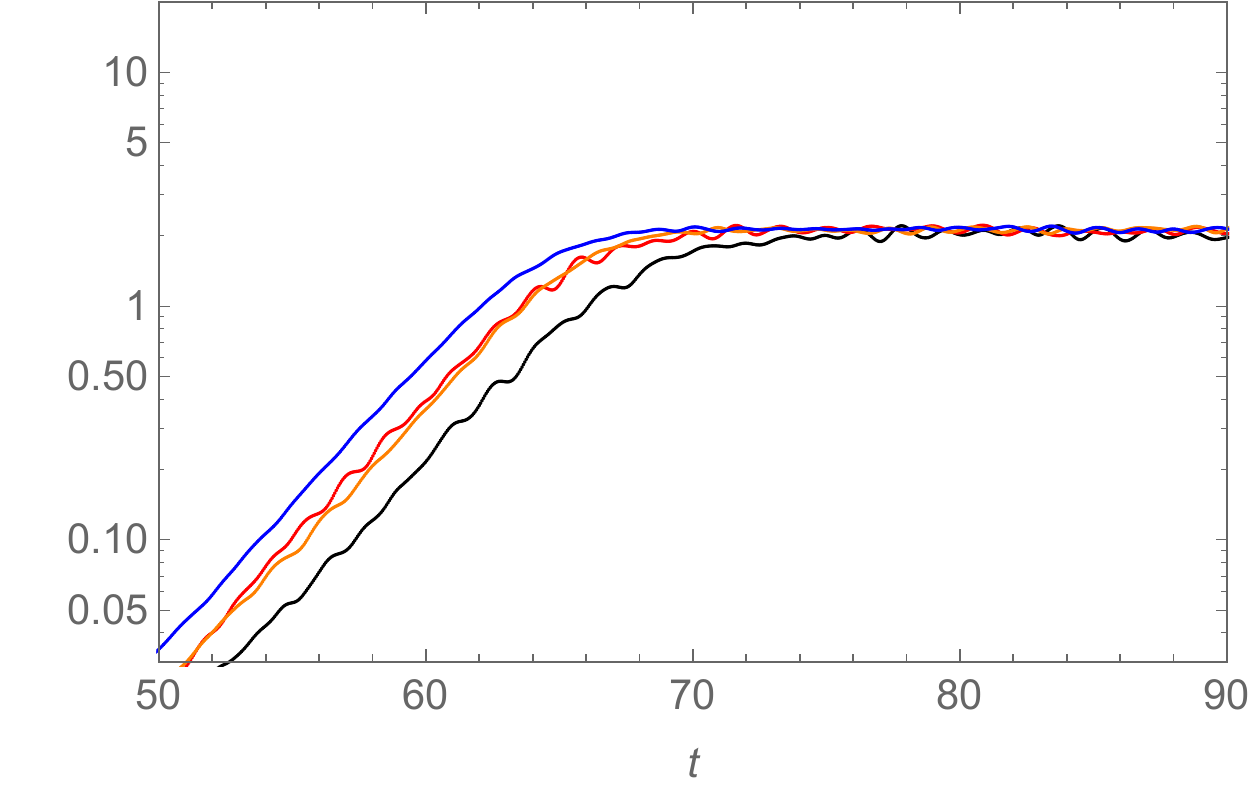}
  \caption{Last phase of the evolution of $|\dX(t)|/\sqrt{N}$ for $N=6$ (black) through $N=16$ (blue). At long times, $|\dX(t)|$ converges to a value that scales as $\sqrt{N}$.}
  \label{fig:finalDx}
\end{figure}

Let us now show that our classical system is a `fast scrambler', namely one in which the scrambling time $t_*$ scales as $\log N^2$.
The typical value of $|\dX(t)|$ when it stops growing can be estimated by picking two random configurations $X$ and $X'$
from the ensemble and calculating the difference between them, $|X-X'|=\sqrt{{\rm Tr}((X-X')^2)}\sim \sqrt{N}$. 
We therefore expect the scrambling time $t_*$ to be given by
\begin{align}
  e^{\lambda_L t_*} \sim \sqrt{N} \Hence
  t_* \sim \frac{1}{4\lambda_L} \log N^2 \ed
\end{align}
We have already seen that $\lambda_L$ is independent of $N$ to leading order.
It is left to show that the perturbation indeed grows to be of order $\sqrt{N}$.
Fig. \ref{fig:finalDx} shows the late-time evolution of $|\dX|$ for various $N$ values.
One can verify numerically that at late times $|\dX| \sim \sqrt{N}$ as expected to within less than one percent.
This establishes fast scrambling in the classical matrix model.
%%%%%%%%%%%%%%%%%%%%%%%%%%%%%%%%%%%%%%%%
%%%%%%%%%%%%%%%%%%%%%%%%%%%%%%%%%%%%%%%%
%%%%%%%%%%%%%%%%%%%%%%%%%%%%%%%%%%%%%%%%
\subsection{Lyapunov Exponent from Poisson Brackets}
%%%%%%%%%%%%%%%%%%%%%%%%%%%%%%%%%%%%%%%%
%%%%%%%%%%%%%%%%%%%%%%%%%%%%%%%%%%%%%%%%
%%%%%%%%%%%%%%%%%%%%%%%%%%%%%%%%%%%%%%%%
The calculations described so far were classical in nature, relying on the time evolution of nearby points in phase space.
On the other hand, computations of the Lyapunov exponent in quantum systems rely on 
commutators and out-of-time-order correlation functions \cite{Larkin1969,Almheiri:2013hfa}.
In this section we bridge this gap by extracting the exponent from the classical limit of commutators --- Poisson brackets.
The results agree with the ones obtained using the previous method.

To motivate the method from a classical perspective, consider a particle in $D$-dimensional space with spatial coordinates $x_I$ and momenta $\pi_I$, where $I=1,\dots,D$.
One can use the classical correlator\footnote{
Classical correlators are defined by time-averaging (assuming ergodicity):
\begin{align}
  \left< O_1(t_1) \cdots O_n(t_n) \right> = 
  \lim_{\tau \to \infty} \frac{1}{\tau} \int^{\tau}_0 
  \! dt' \, O_1(t_1+t') \cdots O_n(t_n+t') \ed
\end{align}
}
\begin{align}
  \left< \{ x_I(t), \pi_J(0) \}_{\pb}^2 \right> = 
  \left< \left( \frac{\dho x_I(t)}{\dho x_J(0)} \right)^2 \right> \sim
  e^{2 \lambda_L t} \ecq
  I \ne J \ec
  \label{xp}
\end{align}
to give an equivalent definition of the Lyapunov exponent $\lambda_L$ \cite{Larkin1969}.
Here we take $I \ne J$ to ensure that the 1-point function vanishes.\footnote{
The averaging in the one-point function $\left< \{ x_I(t), \pi_J(0) \}_{\pb} \right>$ may exhibit large fluctuations that would spoil a clean exponential growth.
We would therefore like to ensure that the `disconnected part' $\left< \{ x_I(t), \pi_J(0) \}_{\pb} \right>^2$ of \eqref{xp} vanishes.
}
We expect that correlators of the form $\left< \{ V(t), W(0) \}_{\pb}^2 \right>$ (where $V,W$ are operators that are local in time) exhibit the same exponential growth as \eqref{xp}.

In the matrix model we would like to focus on gauge-invariant correlators that have a good large $N$ limit.
We will first consider the correlator $\left< O^{ij}(t,0)^2 \right>$ (with no summation over $i,j$), where
\begin{align}
  O^{ij}(t,0) &= 
  \frac{1}{2}
  \left\{ \trace(X^i(t) X^j(t)) , \trace(\Pi^k(0) \Pi^k(0)) \right\}_{\pb} \cr
  &= 
  \Pi^k_{ba}(0) \Bigg[
  \frac{\dho X^{i}_{c d}(t)}{\dho X^k_{ab}(0)}
  X^{j}_{d c}(t) +
  \frac{\dho X^{j}_{c d}(t)}{\dho X^k_{ab}(0)} X^{i}_{d c}(t)
  \Bigg] \ed
  \label{Oij}
\end{align}
Here $\Pi^i$ is the conjugate momentum to $X^i$.
We set $i \ne j$ so that the one-point functions $\langle O^{ij}(t,0) \rangle$ vanish by $SO(9)$ symmetry.
The growth of the correlator is driven by the derivatives in \eqref{Oij}, which are analogous to the derivative in \eqref{xp}.
We therefore expect the correlator to grow as
\begin{align}
  \langle O^{ij}(t,0)^2 \rangle \sim e^{2\lambda_L t} \ec
  \label{Oijcorr}
\end{align}
where $\lambda_L$ is the Lyapunov exponent of the matrix model.

Computing the correlator consists of the following steps.
First, thermalize the system as before by evolving a random initial configuration for time $t_0 = 4000$ to obtain a reference configuration $(X,V)$.
Next, define the perturbed configuration $(X',V) = (X+\dX,V)$ where $\dX^i$ is a polynomial in $V^i$ with small, random coefficients.
Given the reference configuration $(X,V)$ and the perturbed configuration $(X+\dX,V)$, evolve both in time and compute $O^{ij}(t)^2$ (the derivatives in \eqref{Oij} are approximated by replacing $\dho X(t) \to X'(t) - X(t)$).
Finally, average the results over different choices of $i \ne j$ (which are related by $SO(9)$ symmetry), as well as over different initial values and choices of perturbation.

\begin{figure}
  \centering
  \includegraphics[width=0.7\textwidth]{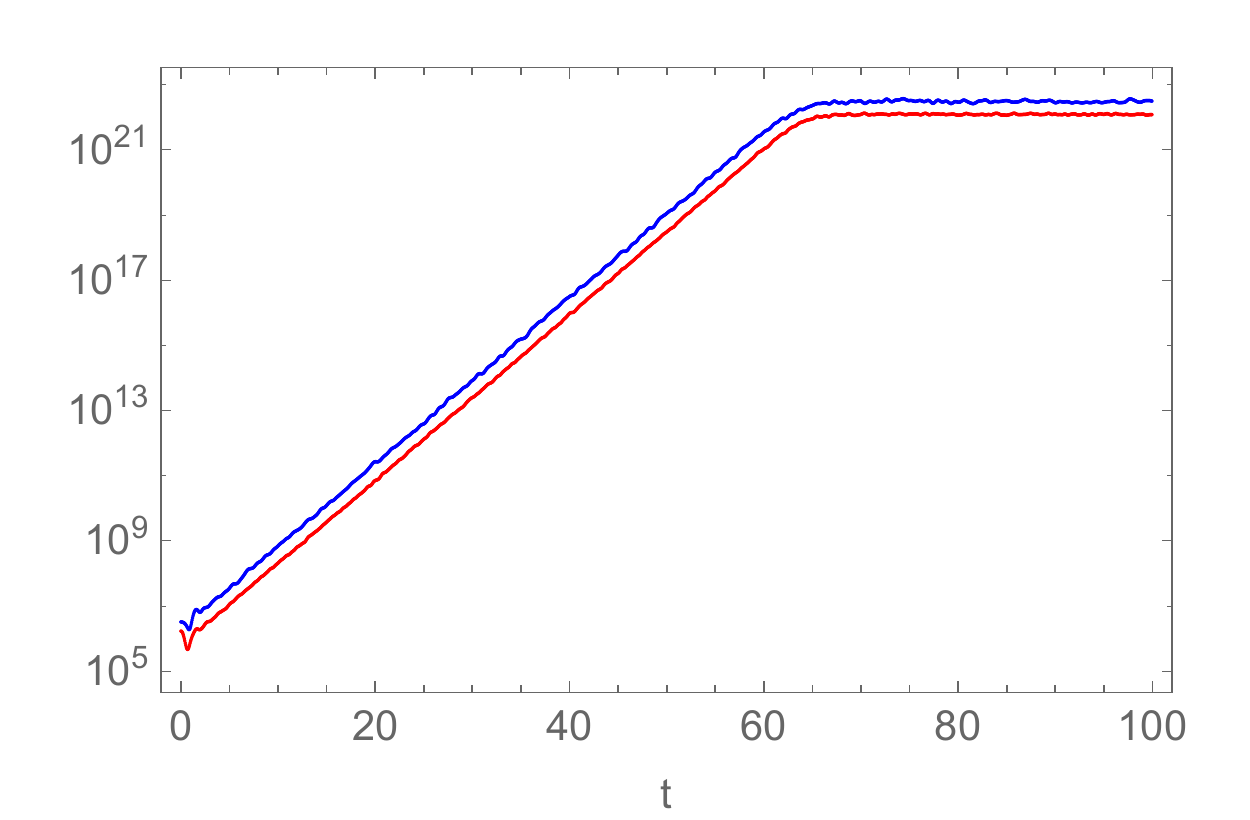}
  \caption{The correlators $\left< O^{ij}(t,0)^2 \right>$ (blue) and $\left< O^{ijk}(t,0)^2 \right>$ (red) as a function of time, with $N=20$. 
  $t=0$ is the time of the initial perturbation.
  }
  \label{fig:pb2pb3}
\end{figure}

\begin{figure}
  \centering
  \includegraphics[width=0.6\textwidth]{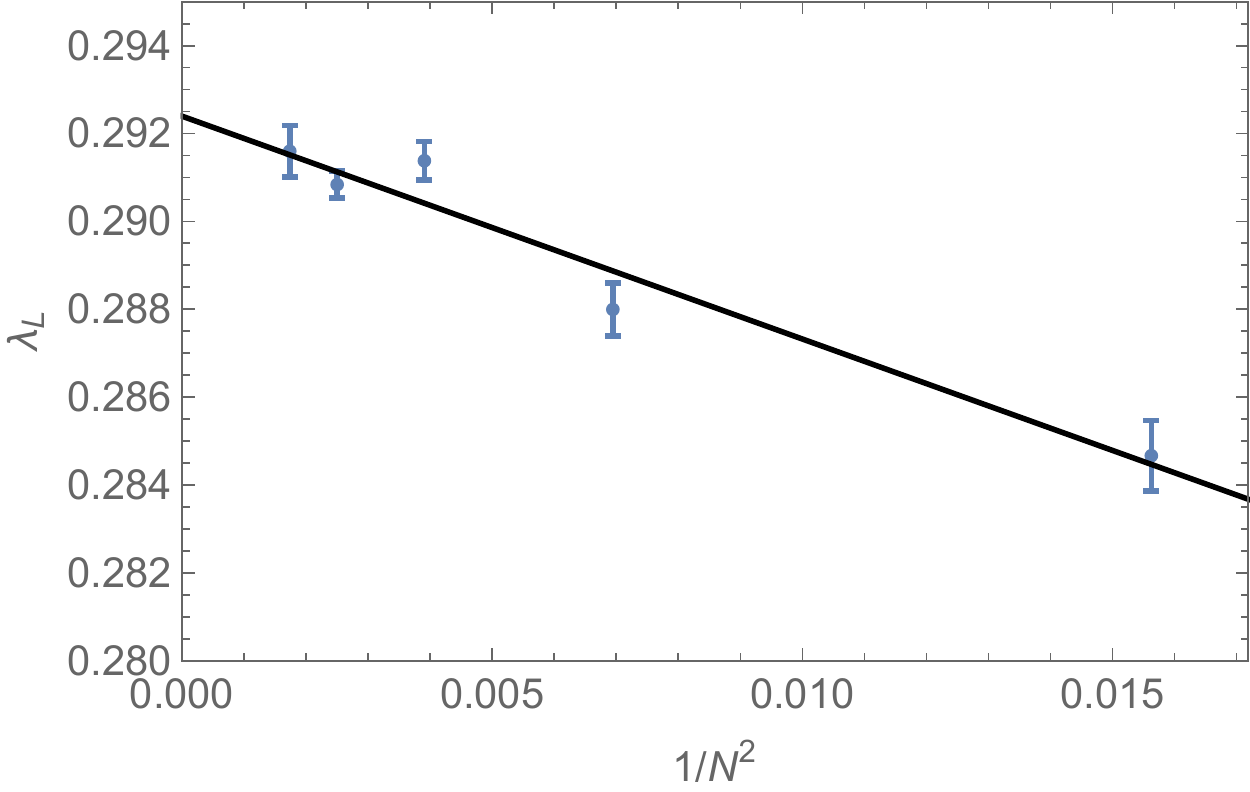}
  \caption{Lyapunov exponents obtained by fitting the growing part of $\langle O^{ij}(t,0)^2 \rangle$, as a function of $1/N^2$.
  The solid line corresponds to the fit \eqref{pbfit}.
  }
  \label{fig:pb2exp}
\end{figure}

The resulting correlator \eqref{Oijcorr} is shown in Fig.~\ref{fig:pb2pb3}.
An initial transient is followed by exponential growth, which saturates when the distance between the reference and perturbed configurations becomes of the same order as the system size.
The fact that the growth stops at $t \simeq 60$ is an artifact of our approximation of the derivative in \eqref{Oij} using finite distances; the exact correlator keeps growing indefinitely.
Fig.~\ref{fig:pb2exp} shows the Lyapunov exponents we get by fitting the growing part of the curves.\footnote{
For each sample of $\left< O^{ij}(t,0)^2 \right>$, we average the values at each $t$ over $i \ne j$ and then fit an exponent.
The fitting window is between $10^{-3}$ and $10^{-11}$ times the saturation (late time) value of the correlator in the averaged sample.
We then average the exponents that are obtained in this way from a few tens of samples with given $N$ value.
The error bars in Fig.~\ref{fig:pb2exp} denote the statistical errors from this averaging of exponents.
}
The large $N$ behavior is given by\footnote{
As in \eqref{lambda}, the uncertainties quoted here do not take into account the error bars.
}
\begin{align}
  \lambda_L = \left[ 0.2924(3)
  - \frac{0.51(6)}{N^2} \right]
  \lambda_{\rm eff}^{1/4} \, T \ed
  \label{pbfit}
\end{align}
This is consistent with the previous result \eqref{lambda} obtained using Sprott's algorithm.

As mentioned above, we expect the Lyapunov exponent to not depend on the operators we use in the Poisson brackets.
To test this, consider the correlator $\left< O^{ijk}(t,0)^2 \right>$ where
\begin{align}
  O^{ijk}(t,0) &= 
  \frac{1}{2}
  \left\{ \trace(X^i(t) X^j(t) X^k(t)) , \trace(\Pi^l(0) \Pi^l(0)) \right\}_{\pb} 
  \ed
  \label{Oijk}
\end{align}
The 1-point function of $O^{ijk}$ vanishes for any choice of $i,j,k$.
The result is shown in Fig.~\ref{fig:pb2pb3}, and the Lyapunov exponent we obtain from this correlator is consistent with the previous results.

%%%%%%%%%%%%%%%%%%%%%%%%%%%%%%%%%%%%%%%%
%%%%%%%%%%%%%%%%%%%%%%%%%%%%%%%%%%%%%%%%
%%%%%%%%%%%%%%%%%%%%%%%%%%%%%%%%%%%%%%%%
\section{Lyapunov Spectrum Computation}
\label{sec:LyapunovSpectrum}
%%%%%%%%%%%%%%%%%%%%%%%%%%%%%%%%%%%%%%%%
%%%%%%%%%%%%%%%%%%%%%%%%%%%%%%%%%%%%%%%%
%%%%%%%%%%%%%%%%%%%%%%%%%%%%%%%%%%%%%%%%

In this section we go beyond the largest exponent and study the full spectrum of Lyapunov exponents \cite{oseledets1968multiplicative}, as defined in Sec.~\ref{lyapunov}.
The evolution of a perturbation $\dX^i,\dV^i$ is given by the linearization of the equations of motion \eqref{eom2}.
Explicitly,
\begin{align}
  \delta \dot{X}^i = \dV^i \ecq
  \delta \dot{V}^i = M^i_{~j}(x) \dX^j \ec
\end{align}
where
\begin{align}
  (M(x) \, \dX)^i \equiv [ \dX^j, [X^i, X^j]] +
  [X^j, [\dX^i, X^j]] + 
  [X^j, [X^i, \dX^j]] \ed
\end{align}
After discretization, the perturbation evolves according to
\begin{align}
  \begin{pmatrix}
    \dX(t+\dt) \\
    \dV(t + \dt)
  \end{pmatrix} =
  U(\dt;x(t))
  \begin{pmatrix}
    \dX(t) \\
    \dV(t)
  \end{pmatrix} \ec
\end{align}
where $U(\dt; x(t))$ is the transfer matrix for a single time step. 
Our discretization \eqref{disc} preserves the Gauss law constraint, and therefore the discretized transfer matrix should preserve the linearized constraint:
\begin{align}
  G_{(X,V)} \begin{pmatrix}
    \dX \\ \dV
  \end{pmatrix} \equiv \sum_{i=1}^d \left( 
  [\dX^i,V^i] + [X^i,\dV^i]
  \right) = 0 \ec
  \label{linear_gauss}
\end{align}
where $d=9$ is the number of matrices. 
The transfer matrix that has this property is
\begin{align}
U(\dt;x(t))
=
\left(
\begin{array}{cc}
1 + \frac{(\delta t)^2}{2}{M}(t)& \delta t\\
 \frac{\delta t}{2}(M(t) + M(t+\delta t))
+ \frac{(\delta t)^3}{4}{M}(t+\delta t){M}(t)
 & 1+\frac{(\delta t)^2}{2}{M}(t+\delta t)
\end{array}
\right).
\nonumber\\
\end{align}
Here we use the notation $M(t) \equiv M(x(t))$.
Composing the single-step transfer matrices gives the finite time transfer matrix $U(t;x_0)$, which evolves a perturbation according to
\begin{align}
  \begin{pmatrix}
    \dX(t) \\
    \dV(t)
  \end{pmatrix} = U(t;x_0)
  \begin{pmatrix}
    \dX \\
    \dV
  \end{pmatrix} \ed
\end{align}

\begin{figure}
  \centering
  \includegraphics[width=0.8\textwidth]{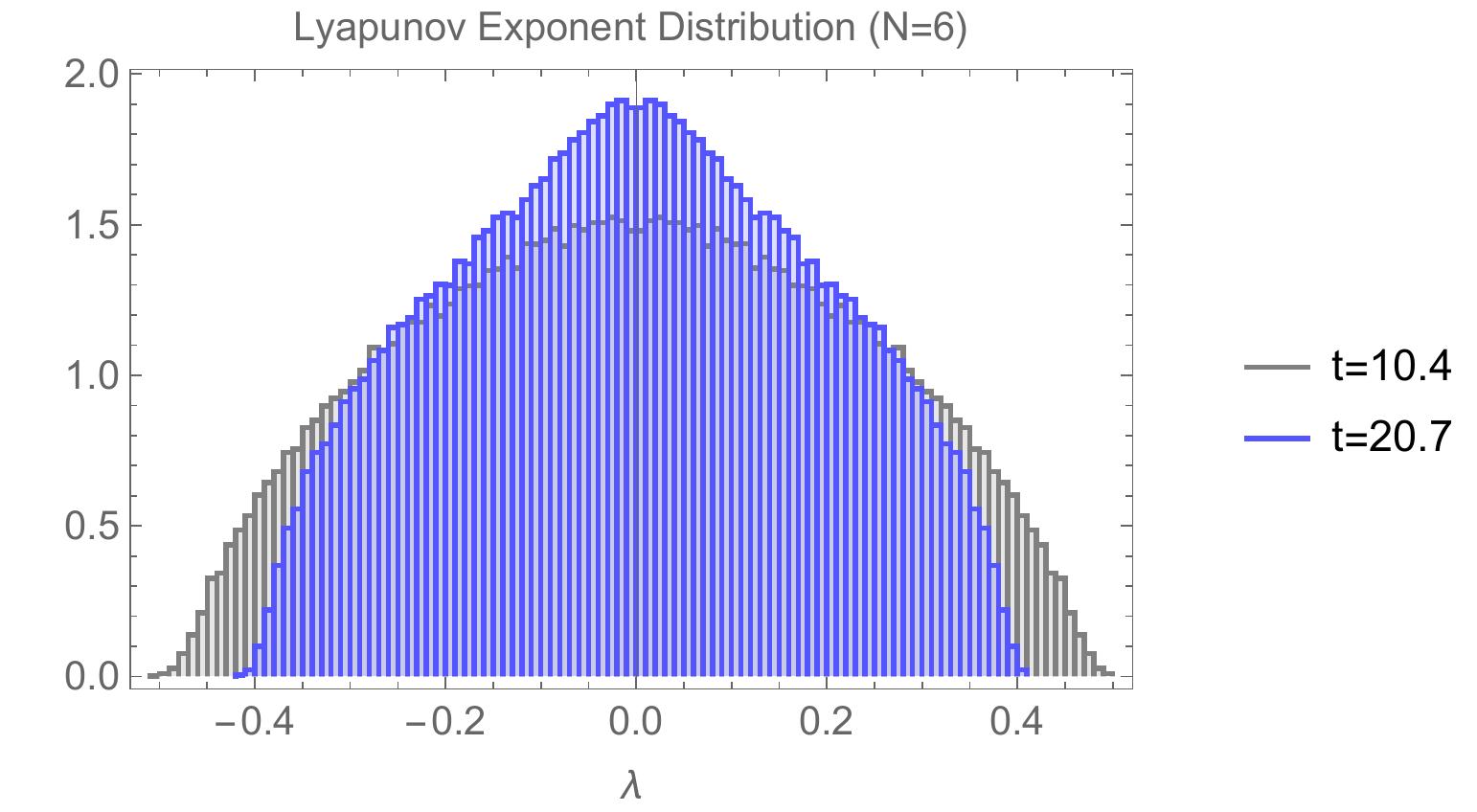}
  \caption{Distribution of Lyapunov exponents with $N=6$ at different times, both normalized with unit area.}
  \label{fig:N6spec}
\end{figure}

\begin{figure}
  \centering
  \includegraphics[width=0.6\textwidth]{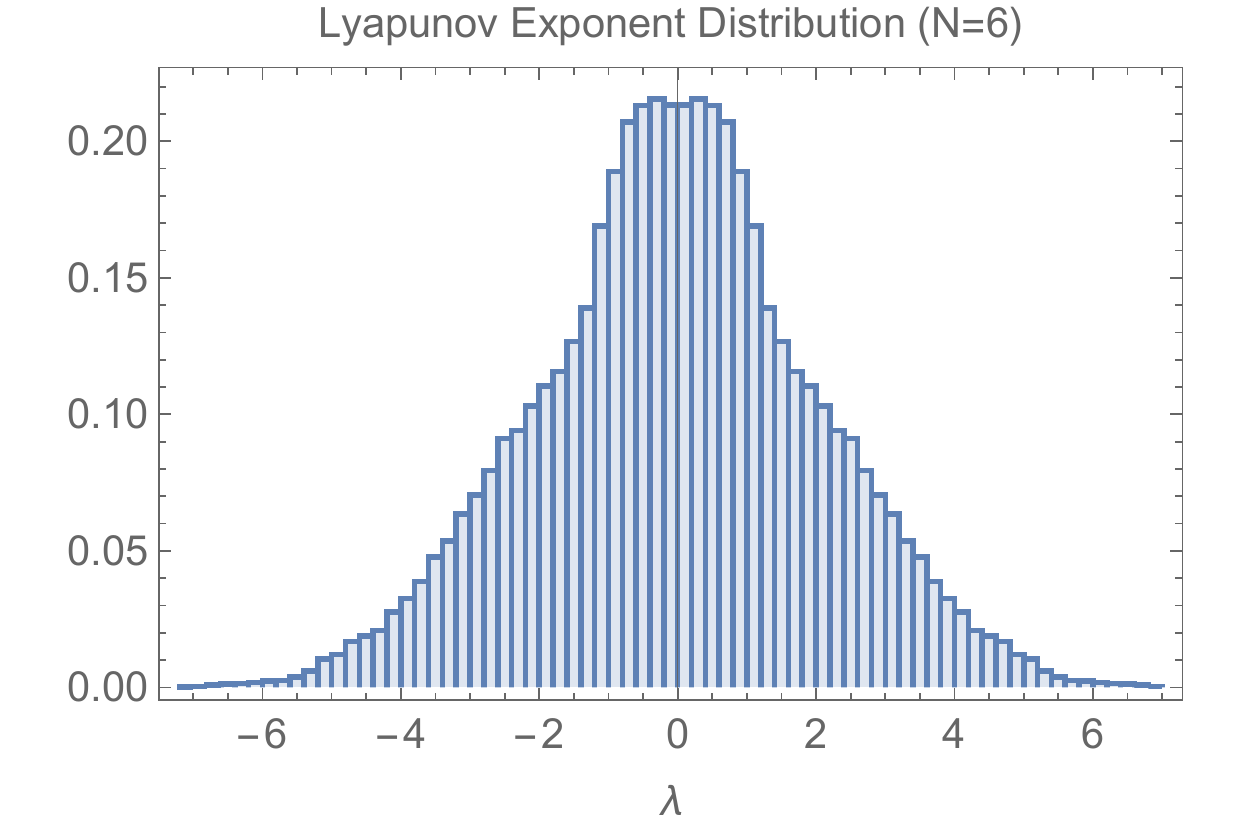}
  \caption{The Lyapunov spectrum for $N=6$ at $t=\dt$, averaged over 60 samples.}
  \label{fig:LocalSpectrum}
\end{figure}

The Lyapunov spectrum can be obtained from the singular values of the transfer matrix, by properly removing unphysical modes. 
As explained in Appendix~\ref{lyapapp}, to obtain the physical exponents we must project out perturbations that do not obey the Gauss law constraint, as well as perturbations that correspond to residual gauge transformations.
For the purpose of computing the spectrum, we find it convenient to work in the linear space $(X,V)$ of $2d$ Hermitian $N \times N$ matrices.
Therefore, we will also include an explicit orthogonal projector onto traceless Hermitian matrices.
In the following, we construct three orthogonal projectors with respect to the metric \eqref{g}.
\begin{enumerate}
\item 
  $P_{U(1)}(x)$ projects out the decoupled $U(1)$ vectors.
\item
  $P_{\rm Gauss}(x)$ projects out vectors at $x=(X,V)$ that do not satisfy the Gauss law condition \eqref{linear_gauss}.
  In particular, it projects onto $\ker(G_{x})$.
  To do this, consider the subspace at $x=(X,V)$ that is spanned by the vectors
  \begin{align}
    \begin{pmatrix}
      [V^i,T^a]\\
      -[X^i,T^a]
    \end{pmatrix}
    \ec \label{gauss-space}
  \end{align}
  where $T^a$ are anti-Hermitian $SU(N)$ generators.
  Notice that the condition \eqref{linear_gauss} is equivalent to saying that a given perturbation $(\dX,\dV)$ is orthogonal to this subspace.
  To define the projector, we can therefore construct an orthonormal basis $\{\vec{v}_a\}$ of the subspace, and write $P_{\rm Gauss}(t)\equiv 1-\sum_a \vec{v}_a\cdot \vec{v}_a^\dagger$.
\item
  $P_{\rm gauge}(x)$ project out pure gauge modes. 
  The pure gauge modes at $x = (X,V)$ are spanned by the vectors
  \begin{align}
    \begin{pmatrix}
      [X^i,T^a]\\
      [V^i,T^a]
    \end{pmatrix}
    \ed \label{gauge-space}
  \end{align}
  By using an orthonormal basis of this space $\{\vec{w}_a\}$, 
  the projector can be defined as $P_{\rm gauge}(t)\equiv 1-\sum_a \vec{w}_a\cdot \vec{w}_a^\dagger$. 
\end{enumerate}
We now define $P(x)\equiv P_{\rm gauge}(x)\cdot P_{\rm Gauss}(x)\cdot P_{U(1)}(x)$. 
It is easy to verify that $P(x)$ is an orthogonal projector.\footnote{
To see this, note that the subspaces spanned by \eqref{gauss-space} and \eqref{gauge-space} are orthogonal when $x$ obeys the Gauss law constraint.
It then follows that $P_{\rm Gauss}(x)$ commutes with $P_{\rm gauge}(x)$.
}
The physical transfer matrix is then defined by (c.f. \eqref{Uphys})
\begin{align}
  U_{\phys}(t;x_0) \equiv P(x(t))\cdot U(t;x_0)\cdot P(x_0) \ed
\end{align} 
This physical transfer matrix has $n = 2(d-1)(N^2-1)$ nonzero singular values, and the $n$ physical Lyapunov exponents can be computed from these by using \eqref{lambdai}.
Fig.~\ref{fig:N6spec} shows the spectrum of the time-dependent exponents \eqref{tlambda} for $N=6$. Numerics\footnote{At longer times the pairing between positive and negative exponents breaks down, and we no longer trust the numerical calculation.
At larger $N$ values the breakdown occurs before $t=20$.
One could improve this by reducing the simulation time step $\dt$, because the numerical violation of the symplectic nature of the transfer matrix grows $\dt \cdot t$.
}
limit us to this modest value of $N$.  But the rapid convergence to the infinite $N$ limit displayed above indicates that these results will be meaningful.
Notice that the largest exponent is larger than the Lyapunov exponent $\lambda_L^{(N=6)} \simeq 0.28$, and that it decreases with time.
In Fig.~\ref{fig:LocalSpectrum}, the spectrum at a single time step $t = \dt$ is shown. The largest exponent is larger by an order of magnitude compared to $t\to\infty$. 
What causes this suppression?
Consider the singular vector $v(t)$ of $U_{\rm phys}(t;x_0)$ that corresponds to the maximal singular value.
If $v(t)$ stayed roughly constant, then the perturbation $\dx$ would quickly align itself with $v(t)$, and the Lyapunov exponent would correspond to the maximal short-time exponent.
Instead, our numerical results suggest that $v(t)$ evolves quickly in time, such that the perturbation cannot become aligned with it.
This suppresses the exponent over time, leading to a smaller $\lambda_L$.

\begin{figure}[t!]
  \centering
  \begin{subfigure}[b]{0.9\textwidth}
    \centering
    \includegraphics[width=\textwidth]{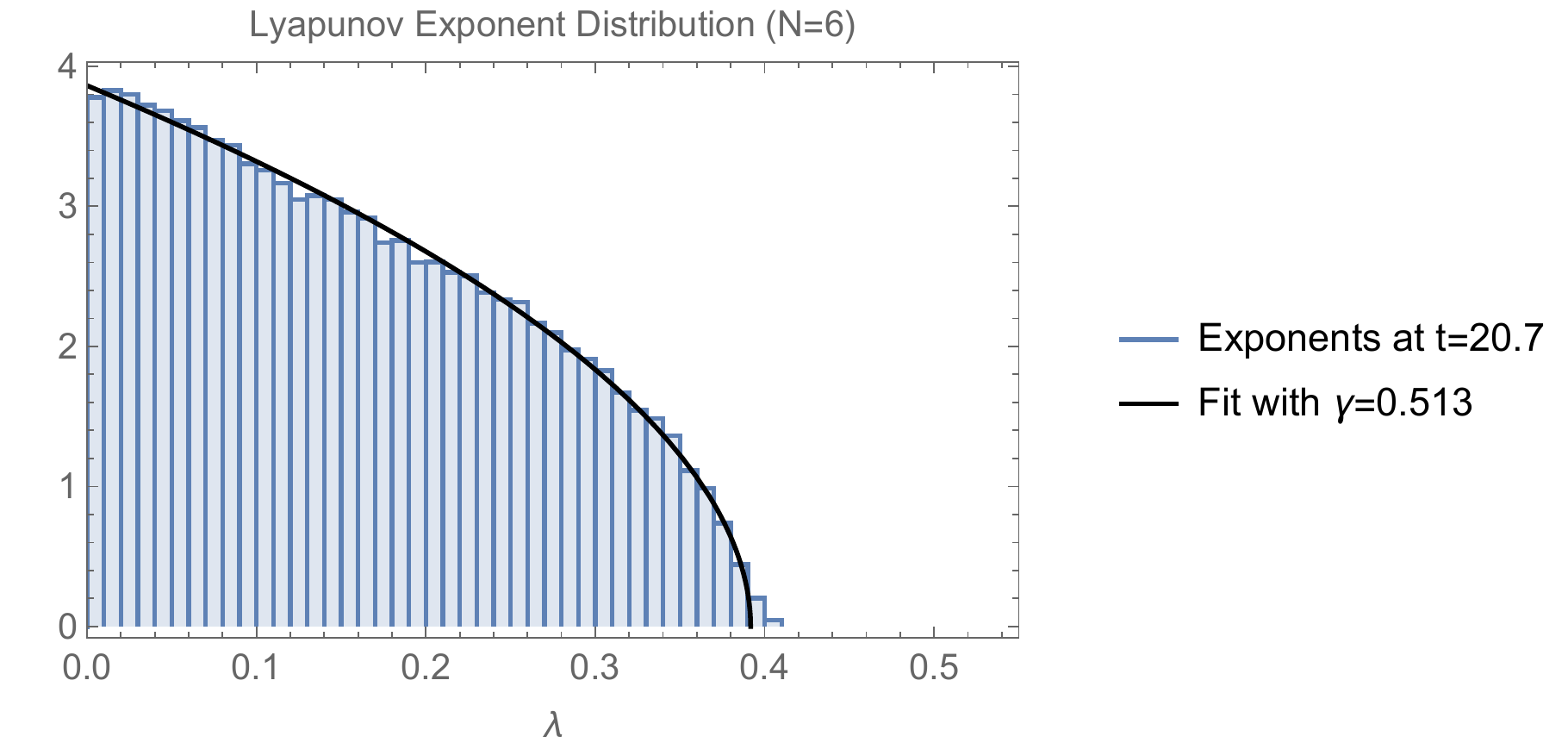}
    \caption{}
  \end{subfigure}
  \begin{subfigure}[b]{0.6\textwidth}
    \centering
    \includegraphics[width=\textwidth]{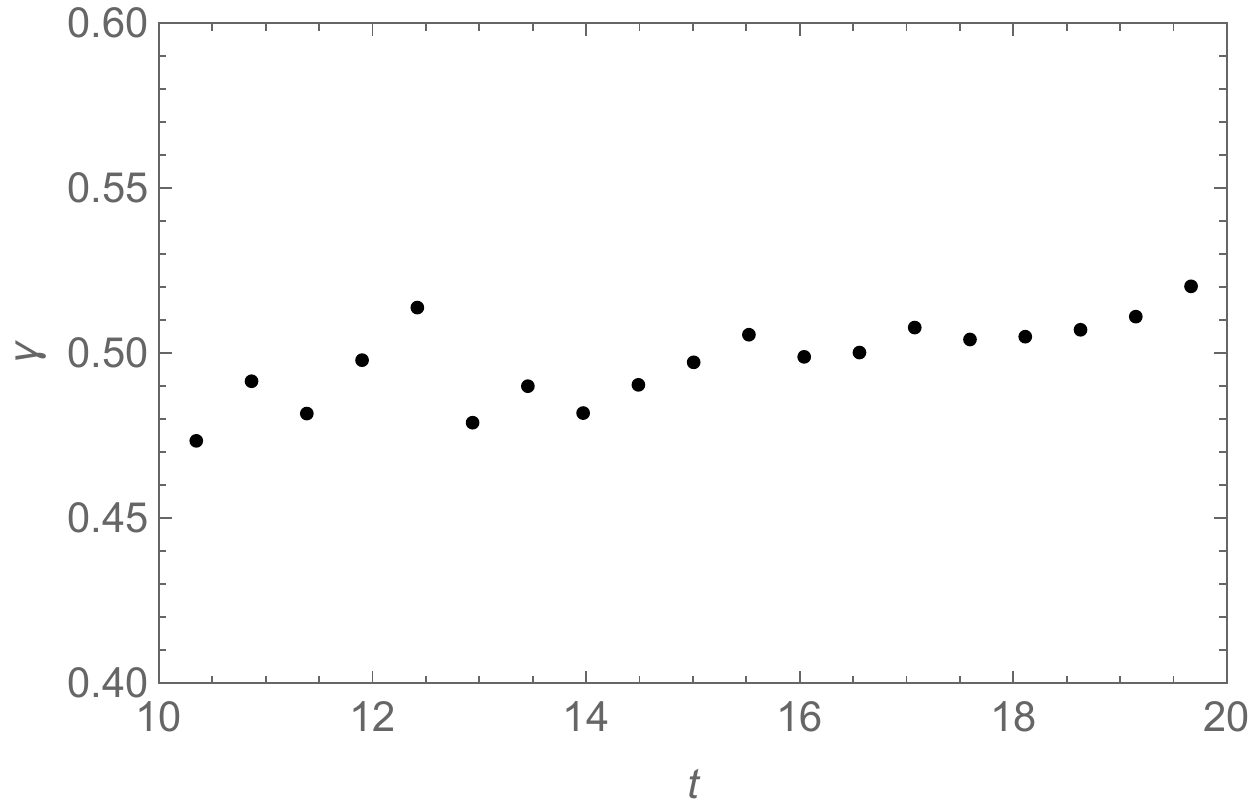}
    \caption{}
  \end{subfigure}
   \caption{(a) Positive Lyapunov spectrum for $N=6$ and a fit to the
   ansatz \eqref{fit_ansatz} at the largest $t$ we have studied, and (b) the fitting parameters $\gamma$ versus $t$.
   Here we normalize the area of only the positive part of the spectrum to unity, and multiply the right-hand side of \eqref{fit_ansatz} by 2 accordingly.
   }\label{fig:GlobalSpec_N6_fit}
\end{figure}

At $t\gtrsim 10$, the spectrum is well described by the ansatz
\begin{eqnarray}
  \rho(\lambda,t)=\frac{(\gamma+1)}{2\tilde{\lambda}_{\rm max}^{\gamma+1}} (\tilde{\lambda}_{\rm max}-|\lambda|)^\gamma \ec \label{fit_ansatz}
\end{eqnarray}
where $\tilde{\lambda}_{\rm max}$ and $\gamma$ both depend on time.
Fig.~\ref{fig:GlobalSpec_N6_fit} shows the finite-time positive Lyapunov spectrum for $N=6$ and a fit to the ansatz \eqref{fit_ansatz}. 
(Note that this $\tilde{\lambda}_{\rm max}$ is a fitting parameter and is not exactly the same value as the largest Lyapunov exponent 
measured in the simulation.) 
We can see that $\tilde{\lambda}_{\rm max}$ decreases with $t$ (see also Fig.~\ref{fig:N6spec}), while $\gamma$ is consistently close to $0.5$. 
More generally, we found that $\gamma = 0.5 \pm 0.1$ in all checks we made.

There are two exponents at finite $t$ which should both converge to the Lyapunov exponent as $t\to\infty$: 
The largest exponent determined from the transfer matrix, which we call $\lambda_{\rm max}(t)$, 
and the `effective' exponent calculated in Sec.~\ref{sec:leading_exponent}, which is defined by 
\begin{align}
  |\dx(t)|^2 = e^{2\lambda_L(t) t} |\dx(0)|^2 \ed 
\end{align}
As shown in Sec.~\ref{sec:syserr}, for generic perturbations we can approximate this exponent by
\begin{align}
  \lambda_L(t) &\simeq 
  \frac{1}{2t} \log \left( 
  \frac{1}{n} \sum_i e^{2\lambda_i(t) t} \right) 
  \ed
  \label{lLt}
\end{align}

\begin{figure}[t!]
  \centering
  \includegraphics[width=\textwidth]{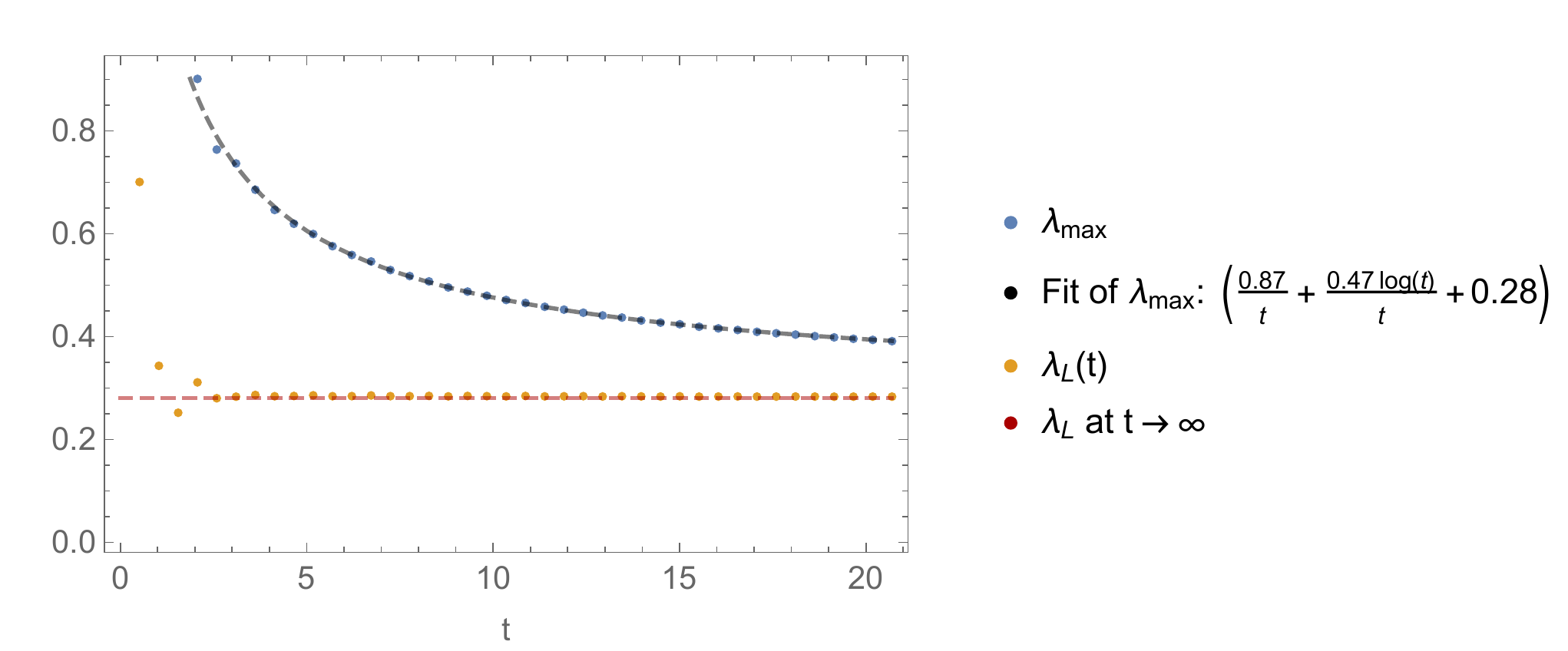}
  \caption{ 
  The largest and average exponents as a function of time, with $N=6$. 
  Data is an average over 80 samples.
  }
  \label{fig:exponents_N6}
\end{figure}
Fig.~\ref{fig:exponents_N6} compares these exponents.
It is surprising that $\lambda_L(t)$ quickly approaches its asymptotic value and then remains essentially constant, while $\lambda_{\rm max}(t)$ converges much more slowly.
We do not have an explanation for this behavior.
It is consistent with the clean exponential growth of perturbations that we observed in Sec.~\ref{largest}.
We tried several fitting ansatz to match the evolution of $\lambda_{\rm max}(t) $ such as $a + \frac{b}{t}$, $a + \frac{b}{t}+\frac{c}{t^2}$, 
and $a + \frac{b}{t^c}$. 
It turned out $a + \frac{b}{t} + \frac{c\log t}{t}$ fits the data very well at a wide time window, 
and attains the correct late-time value $\lambda_{\rm max}(t=\infty)\simeq 0.28$ determined in Sec.~\ref{largest}.\footnote{
In fact, we noticed numerically that all exponents, including the maximal one, are well-approximated by the ansatz 
$\lambda_i(t) = \lambda_i + \frac{a_i \log(2\pi t)}{t}$.}
By recalling the fact that $\gamma$ stays close to $0.5$, 
we can expect the late-time behavior to be 
\begin{eqnarray}
\rho(\lambda,t=\infty)=\frac{(\gamma+1)}{2\lambda_L^{\gamma+1}} (\lambda_L-|\lambda|)^\gamma, 
\end{eqnarray}
where $\lambda_L\simeq 0.28$ is the largest Lyapunov exponent determined in Sec.~\ref{largest}
and $\gamma\simeq 0.5$. 
The relation to random matrix theory has not escaped our attention, although we do not see how to make it precise.
It would be interesting to see whether it holds for larger values of $N$ as well.\footnote{
One way of studying the late-time Lyapunov spectrum is by a well-known generalization of the Sprott's algorithm.
To compute the leading $k$ exponents one chooses $k$ initial orthonormal vectors, evolves each one according to Sprott's procedure, and also re-orthogonalizes the vectors after each time step.
}

%%%%%%%%%%%%%%%%%%%%%%%
%%%%%%%%%%%%%%%%%%%%%%%
%%%%%%%%%%%%%%%%%%%%%%%
\section{Discussion}
\label{discussion}
%%%%%%%%%%%%%%%%%%%%%%%
%%%%%%%%%%%%%%%%%%%%%%%
%%%%%%%%%%%%%%%%%%%%%%%
The above results show that the  Lyapunov spectrum approaches a smooth $N$-independent limit as $N\rightarrow \infty$ in this classical system. This is consistent with the observation of Sekino and Susskind \cite{Sekino:2008he} about the $k$-locality of the matrix quantum mechanics Hamiltonian in matrix index space.  

In addition, these results bring into focus the existence of a whole spectrum of Lyapunov exponents in these large $N$ systems.   This raises the question about the nature of the corresponding spectrum in  quantum large $N$ systems and their meaning in the bulk gravity dual.   There is one indication of this in existing work:  In \cite{Shenker:2014cwa} it was pointed out that stringy corrections to Einstein gravity modify the rate of exponential growth of scrambling, tending to decrease it.  In particular the result found there is
\begin{align}
\lambda_L(k_T) = \frac{2 \pi}{\beta}\left( 1 - \frac{c_1}{\sqrt{\lambda}} - \frac{c_2}{\sqrt{\lambda} }k_T^2 \right)
\end{align}
Here $k_T$ is the transverse momentum of the perturbation in the field theory space and $c_1$ and $c_2$ are known positive constants.   This gives a spectrum of Lyapunov exponents indexed by $k_T$.  That chaos develops at different rates for different modes gives a physical interpretation of the diffusion of chaos from a localized perturbation found in \cite{Shenker:2014cwa}.   The analog of $k_T$ and its role in the spectrum in the setup described in this paper  where there is not a field theoretic transverse space are interesting questions for future work.

Another important question is the  quantum field theory and holographic origin of the order  $N^2$ different Lyapunov  exponents found in the classical system.  Weakly coupled field theory analyses \cite{Shenker:2014cwa,Stanford_talk_2015} should shed some light on the QFT question.   We hope to return to this in subsequent work.

As discussed above, the existence of a spectrum of Lyapunov exponents immediately brings to mind Pesin's theorem and the Kolmogorov-Sinai entropy one gets by adding up the positive ones.  This raises the important question of the meaning of the KS entropy in large $N$ quantum systems and in their holographic duals.   Suggestions for definitions of quantum KS entropy have been made in the literature \cite{PhysRevA.66.052302,alicki1998quantum,mendes1995entropy,man2000lyapunov,man2002quantum,Kunihiro01032009}. These should certainly be explored further given recent developments.   We would expect any entropy, including the KS entropy, to be proportional to $N^2$, sharpening the question mentioned above.  

A simple model connecting KS entropy and entanglement entropy has been constructed in \cite{Asplund:2015osa}, motivated by \cite{Zurek:1994wd}. Building on these ideas the authors of \cite{Asplund:2015osa}  have recently constructed a toy field theory model relating entanglement entropy and KS entropy and conjectured a bound on the rate of growth of entanglement that follows from the bound on Lyapunov exponents \cite{Maldacena:2015waa}.

The question about the holographic role of KS entropy and its relation to rates of increase of various entropies has been raised by a number of authors 
\cite{0264-9381-25-19-195005,Balasubramanian:2011ur,Li2013}.  Entanglement entropy growth is a natural candidate that has been discussed in detail \cite{Balasubramanian:2011ur}. One hint of such a connection is an observation of Stanford \cite{stanfordUnpub}. In Einstein gravity the butterfly velocity $v_B$ describing the transverse spread of chaos \cite{Shenker:2013pqa,Roberts:2014isa} is the same as the saturation velocity\footnote{
See equation (11) in \cite{Liu:2013iza}.
}
in the entanglement tsunami picture of \cite{Liu:2013iza} that describes the rate of growth of the spatial  region where entanglement entropy reaches its saturated value. This connection occurs because the Ryu-Takayanagi surface computing the entanglement entropy in this regime dips very close to the horizon, the region where the exponential blueshifts responsible for holographic chaos occur.

\section*{Acknowledgements} 
\hspace{0.51cm}
%%%%%%%%%%%%%%%%%%%%%%%%%%%%%%%%%%%%%%%%%%%%%%%%%%%%%%%%%%%%%%%%%%%%
%%%%%%%%%%%%%%%%%%%%%%%%%%%%%%%%%%%%%%%%%%%%%%%%%%%%%%%%%%%%%%%%%%%%
%%%%%%%%%%%%%%%%%%%%%%%%%%%%%%%%%%%%%%%%%%%%%%%%%%%%%%%%%%%%%%%%%%%%
The authors would like to thank Ahmed Almheiri, David Berenstein, Micha Berkooz, Evan Berkowitz, Patrick Hayden, Paul Romatschke and Douglas Stanford for stimulating discussions and comments.

The work of M.~H. is supported in part by the Grant-in-Aid of the Japanese Ministry 
of Education, Sciences and Technology, Sports and Culture (MEXT) 
for Scientific Research (No. 25287046). 
The work of G.~G. is supported by a grant from the John Templeton Foundation. 
The work of S.~S. is supported by NSF grant  PHY-1316699 and a grant from the Templeton Foundation.
The opinions expressed in this publication are those of the authors and do not necessarily reflect the views of the John Templeton Foundation.

\appendix

%%%%%%%%%%%%%%%%%%%%%%%
%%%%%%%%%%%%%%%%%%%%%%%
%%%%%%%%%%%%%%%%%%%%%%%
\section{Metric Independence of Lyapunov Exponents}
\label{indep}
%%%%%%%%%%%%%%%%%%%%%%%
%%%%%%%%%%%%%%%%%%%%%%%
%%%%%%%%%%%%%%%%%%%%%%%
In this section we prove that the Lyapunov exponents on a compact Riemannian manifold $(\cM,g)$ are independent of the choice of metric.
A similar argument for the invariance of Lyapunov exponents under coordinate transformations was introduced in \cite{eichhorn2001transformation}.
Let $x(t)$, $t \ge 0$, be a reference trajectory, and let $\dx(t)$ be a tangent vector at $x(t)$ that represents an evolving infinitesimal perturbation.
The Lyapunov exponent for the initial conditions $x_0 = x(0)$, $\delta x_0 = \delta x(0)$ is defined by
\begin{align}
  \lambda_{x_0}(\delta x_0) = \lim_{t \to \infty} \frac{1}{t} 
  \log | \delta x(t) |_g \ec
\end{align}
where $|\delta x|_g = \sqrt{g(\delta x,\delta x)}$.
Now, consider another Riemannian metric $\tilde{g}$ on $\cM$.
The corresponding Lyapunov exponent with respect to this metric is
\begin{align}
  \tilde{\lambda}_{x_0}(\delta x_0) = \lim_{t \to \infty} \frac{1}{t} 
  \log |\delta x(t)|_{\tilde{g}} \ed
\end{align}

Let us show that $\lambda_{x_0}(\delta x_0) = \tilde{\lambda}_{x_0}(\delta x_0)$.
Define $r_+ > 0$ by
\begin{align}
  r_+ &= \max_{x \in \cM} r_+(x) \ecq \cr
  r_+(x) &= \max_{v \in T_x} \frac{|v|_{\tilde{g}}}{|v|_g} 
  = \max_{\begin{smallmatrix} v \in T_x, \\ |v|_g=1 \end{smallmatrix}} |v|_{\tilde{g}}
    \label{r} \ed
\end{align}
Here, $T_x$ is the tangent space at $x$.
The maxima are well-defined because the norm is continuous, and both $\cM$ and the unit ball at each point are compact.
For any $x \in \cM$ and any $v \in T_x$, we then have $|v|_{\tilde{g}} \le r_+ |v|_g$.
Now,
\begin{align}
  \tilde{\lambda}_{x_0}(\delta x_0) =
  \lim_{t \to \infty} \frac{1}{t} \log |\delta x(t)|_{\tilde{g}} \le 
  \lim_{t \to \infty} \frac{1}{t} \log (r_+ |\delta x(t)|_{g}) =
  \lambda_{x_0}(\delta x_0) \ed
\end{align}
The other inequality can be obtained using the definition
\begin{align}
  r_- &= \min_{x \in \cM} \min_{v \in T_x} \frac{|v|_{\tilde{g}}}{|v|_g} \ed
\end{align}
This completes the proof.

%%%%%%%%%%%%%%%%%%%%%%%
%%%%%%%%%%%%%%%%%%%%%%%
%%%%%%%%%%%%%%%%%%%%%%%
\section{Lyapunov Exponents and Gauge Symmetry}
\label{lyapapp}
%%%%%%%%%%%%%%%%%%%%%%%
%%%%%%%%%%%%%%%%%%%%%%%
%%%%%%%%%%%%%%%%%%%%%%%
In this section we construct a `physical' phase space $\cM_{\phys}$ for the matrix model by following the procedure of symplectic reduction (or Marsden-Weinstein reduction).
The physical phase space is free of gauge redundancy.
We then equip the space with a Riemannian metric, which allows us to define gauge-invariant Lyapunov exponents.

To construct the physical phase space we begin with the total space $\cM$, parameterized by $(X,V)$ and equipped with the symplectic form $\omega = \sum dX^i_{ab} \wedge dV^i_{ba}$.
The dimension of $\cM$ is $2d(N^2-1)$,  where $d=9$.
As explained in Sec.~\ref{mme}, gauge redundancy affects this space in two ways.
First, Gauss's law restricts physical configurations to lie on the constrained surface
\begin{align}
  \cM_0 \equiv \Big\{ (X,V) \;\,\Big|\;\, \sum_i [X^i,V^i] = 0 \Big\} \ed
\end{align}
Second, points on $\cM_0$ that are related by a residual gauge transformation \eqref{res} are physically identical.

We define the space $\cM_{\phys}$ by identifying points on $\cM_0$ that are related by a gauge transformation.
The physical phase space is the coset space $\cM_{\phys} \equiv \cM_0 /\!\! \sim$, where for any $x,x' \in \cM_0$ we say that $x \sim x'$ if these points are related by a gauge transformation of the form \eqref{res}.
Points on $\cM_{\phys}$ will be denoted by $[x]$ where $x \in \cM_0$.
$\cM_{\phys}$ will generally have a complicated global structure that includes singularities.
However, a typical point on a given fixed-energy subspace has a smooth neighborhood, and we will only be interested in the local structure at such points.
The dimension of $\cM_{\phys}$ at such points is $2(d-1)(N^2-1)$. 

The tangent space at a point $[x] \in \cM_{\phys}$ is obtained from the tangent space at $x = (X,V) \in \cM_0$ by modding out infinitesimal gauge transformations.
The subspace of gauge transformations at $(X,V)$ is spanned by the vectors $(\dX_{H},\dV_{H})$ where
\begin{align}
  \delta X^i_{H} = i [X^i, H] \ecq
  \delta V^i_{H} = i [V^i, H] \ec
  \label{resinf}
\end{align}
and $H$ is any traceless, Hermitian $N \times N$ matrix.
Vectors on the physical tangent space, which we denote by $[\dx]$, obey $[\dx] = [\dx + (\dX_H,\dV_H)]$ for any Hermitian $H$.

In order to regard $\cM_{\phys}$ as a phase space, we must equip it with a symplectic structure.
Wherever $\cM_{\phys}$ is smooth, we define the natural symplectic form
\begin{align}
  \omega_{\phys}([\dx],[\dx']) = \omega(\dx,\dx') \ed
\end{align}
Here $[\dx],[\dx']$ are vectors at $[x] \in \cM_{\phys}$, and $\dx,\dx'$ are chosen representative vectors at $x$.
It is easy to verify that this definition is independent of the choice of representatives because of the Gauss law constraint. 

To define Lyapunov exponents we must also equip $\cM_{\phys}$ with a Riemannian metric.
Let us use the metric $g$ on $\cM$ (c.f. \eqref{g}) to define a metric $g_{\phys}$ on the physical phase space.
First, restrict $g$ to $\cM_0$ and consider the tangent space at a point $x = (X,V) \in \cM_0$.
Let $P_{\rm gauge}$ denote the orthogonal projection operator (with respect to $g$) that projects out the pure gauge vectors.
We now define the inner product of two vectors $[\dx]=[(\dX,\dV)]$, $[\dx']=[(\dX',\dV')]$ on $\cM_{\phys}$ by
\begin{align}
  g_{\phys}([\dx], [\dx']) \equiv
  g(P_{\rm gauge}(\dX,\dV); P_{\rm gauge}(\dX',\dV')) \ed
  \label{gphys}
\end{align}
On the right-hand side we have chosen representative vectors $(\dX,\dV),(\dX',\dV')$ at $x$.
The metric is well-defined, in that it is independent of the choice of vector representatives and of $x$.
In particular, notice that the problem that prompted the introduction of the physical metric is now solved: Two points on $\cM_0$ that are related by a gauge transformation are equivalent on the reduced $\cM_{\phys}$, and have vanishing distance under the physical metric.

\subsection{Gauge-Invariant Exponents}

Lyapunov exponents can now be defined for fixed energy subspaces of $\cM_{\phys}$ using the physical metric, and they will be independent of our choice of metric as shown in Appendix~\ref{indep}.
The first step is to define a transfer matrix $U_{\phys}$ that only propagates physical modes.
It can be done by a projection 
\begin{align}
  U_{\phys}(t;x_0) \equiv P(x(t)) \cdot U(t;x_0) \cdot P(x_0) \ec
  \label{Uphys}
\end{align}
where $P(x) \equiv P_{\rm gauge}(x) P_{\rm Gauss}(x)P_{U(1)}(x)$ 
is an orthogonal projector defined in Sec.~\ref{sec:LyapunovSpectrum}. 
Given a generic initial vector on $\cM$, 
$P_{\rm Gauss}(x_0)$ restricts the perturbation to lie on $\cM_0$, and 
$P_{\rm gauge}(x_0)$ removes the pure gauge modes. 
This chooses a representative vector on $\cM_{\phys}$.
The vector then propagates with the usual transfer matrix $U(t;x_0)$.
After propagation we project again.

To compute the Lyapunov exponents, perform the singular value decomposition of $U_{\phys}$.
There are $2d(N^2-1)$ singular values, of which $2(N^2-1)$ vanish due to the projections.
The gauge-invariant Lyapunov exponents are computed from the remaining (positive) singular values by using \eqref{lambdai}.

As we now show, the physical transfer matrix $U_{\phys}$ is symplectic with respect to $\omega_{\phys}$. 
As a result, the physical Lyapunov exponents are paired.
To show that $U_{\phys}$ is symplectic, we need to show it obeys the equation
\begin{align}
  U_{\phys}^\dagger(x \to x') \cdot \omega_{\phys}(x') \cdot U_{\phys}(x \to x') = \omega_{\phys}(x)
  \ed \label{Usymp}
\end{align}
Here we introduced the notation $U_{\phys}(t;x) \equiv U_{\phys}(x \to x(t))$ for clarity.
$\omega_{\phys}(x)$ and $\omega_{\phys}(x')$ are matrix representations of $\omega_{\phys}$ using the same bases we use to represent $U_{\phys}$.
They are given by $\omega_{\phys}(x) = P(x) \cdot \omega(x) \cdot P(x)$, where $\omega(x)$ represents the symplectic form on the total phase space (we may choose $\omega(x)$ to be constant, but this is not necessary).
Notice that the matrix $\omega_{\phys}(x)$ generally depends on $x$.
Equation \eqref{Usymp} can be written as
\begin{align}
  P(x) \cdot U^\dagger(x \to x') \cdot P(x') \cdot \omega(x') \cdot P(x') \cdot U(x \to x') \cdot P(x) =
  P(x) \cdot \omega(x) \cdot P(x) \ed
\end{align}
Now we claim that the $P(x')$ factors on the left-hand side re redundant.
To see this, first note that $P(x') U(x \to x') P(x) = P_{\rm gauge}(x') U(x \to x') P(x)$, due to the fact that time evolution preserves the Gauss law constraint.
Further, the remaining factor of $P_{\rm gauge}$ can be dropped because, after reducing to the Gauss-constrained subspace, pure gauge perturbations vanish automatically in the symplectic form.
We therefore are left with the equation
\begin{align}
  P(x) \cdot U^\dagger(x \to x') \cdot \omega(x') \cdot U(x \to x') \cdot P(x) =
  P(x) \cdot \omega(x) \cdot P(x) \ec
\end{align}
which follows immediately from the fact that $U$ is symplectic with respect to $\omega$.
This concludes the proof that $U_{\phys}$ is symplectic on the physical space, and therefore the physical Lyapunov exponents are paired.

%%%%%%%%%%%%%%%%%%%%%%%%%%%%%%%%%%%%%%%
%%%%%%%%%%%%%%%%%%%%%%%%%%%%%%%%%%%%%%%
%%%%%%%%%%%%%%%%%%%%%%%%%%%%%%%%%%%%%%%
\section{Perturbation Compatible with the Gauss Law Constraint}\label{appendix:perturbation}
%%%%%%%%%%%%%%%%%%%%%%%%%%%%%%%%%%%%%%%
%%%%%%%%%%%%%%%%%%%%%%%%%%%%%%%%%%%%%%%
%%%%%%%%%%%%%%%%%%%%%%%%%%%%%%%%%%%%%%%

Given a thermalized configuration $X(t),V(t)$, we would like to perturb it slightly while preserving the Gauss law constraint \eqref{gauss}.
We will do this by deforming the potential energy with additional interaction terms that preserve the constraint, evolving the system for a short time to obtain a perturbed configuration $X'(t),V'(t)$, and then restoring the original Lagrangian.
We add the following term to the potential,
\begin{align}
  \sum_{k=1}^{k_0} c_k {\rm Tr }\left[ \Big( \sum_i X_i^2 \Big)^k \right] \ed
\end{align}
The force is modified from $F^i(t)$ to
\begin{align}
  \tilde{F}^i(t) =
  \sum_{j}[X^j(t),[X^i(t),X^j(t)]]
  +
  \sum_{k=1}^{k_0} kc_k  
  \left\{
  X^i(t) \,, \Big(\sum_j X_j^2(t)\Big)^{k-1}
  \right\} \ec
\end{align}
where $\{ \cdot \,, \cdot \}$ is the anti-commutator.
The Gauss law constraint is still preserved, because the modified force still satisfies the relation $\sum_i [X^i(t), \tilde{F}^i(t)] = 0$.
In practice we choose $k_0 = 2$, the coefficients $c_k$ are chosen randomly from $\cN(0,10^{-8})$, and we evolve the deformed system for time $t_1=1$ before turning off the deformation.\footnote{
Another simple way of a deformation keeping the Gauss's law constraint is an addition of a polynomial of $V_M$ to $X_M$. 
We confirmed that the detail of perturbation does not affect the results. }
%%%%%%%%%%%%%%%%%%%%%%%%%%%%%%%%%%%%%%%
%%%%%%%%%%%%%%%%%%%%%%%%%%%%%%%%%%%%%%%
%%%%%%%%%%%%%%%%%%%%%%%%%%%%%%%%%%%%%%%
\section{Sprott's Algorithm}
\label{sprott}
%%%%%%%%%%%%%%%%%%%%%%%%%%%%%%%%%%%%%%%
%%%%%%%%%%%%%%%%%%%%%%%%%%%%%%%%%%%%%%%
%%%%%%%%%%%%%%%%%%%%%%%%%%%%%%%%%%%%%%%
In this section we describe Sprott's algorithm \cite{sprott2003chaos}, which we use in Sec.~\ref{largest} to compute the leading Lyapunov exponent of the classical matrix model.
The basic idea behind the algorithm is to rescale the perturbation at each time step such that it stays small, the linear approximation continues to hold, and the growth never saturates.
The evolution can then continue until the measured exponent converges with some chosen precision.

\begin{enumerate}
  \item Choose an initial (thermalized) reference point $x_0=(X,V)$ and a perturbed point $x'_0=(X',V')$.
    Let $d_0$ denote the distance between them, computed using our chosen distance function.
  \item At the $n$th iteration, evolve $x_{n-1}$ and $x'_{n-1}$ by one time step $\dt$, obtaining $x_n$ and $\tilde{x}_{n}$ respectively.
    \label{firstStep}
  \item Compute the distance $d_n$ between $x_n$ and $\tilde{x}_n$.
  \item Define the new configuration $x'_n$ by
    \begin{align}
      x'_n = x_n + \frac{d_0}{d_n} (\tilde{x}_n - x_n) \ed
    \end{align}
    The difference has been rescaled such that $x_n$ and $x'_n$ are again at distance $d_0$.
    \label{lastStep}
  \item Repeat steps \ref{firstStep}-\ref{lastStep}.
    The leading Lyapunov exponent is given by
    \begin{align}
      \lim_{n \to \infty}
      \frac{1}{n\dt} \sum_{i=1}^n \log \left( \frac{d_i}{d_0} \right)
      \ed
    \end{align}
\end{enumerate}
Note that the rescaling in step \ref{lastStep} implies that the new configuration $x'_n$ does not satisfy the Gauss law constraint.
However, the violation is subleading in the size of the perturbation, and we verified numerically that the violation remains negligible over the entire time evolution.

%%%%%%%%%%%%%%%%%%%%%%%
%%%%%%%%%%%%%%%%%%%%%%%
%%%%%%%%%%%%%%%%%%%%%%%
\section{Finite Volume of Classical Flat Directions}
\label{appendix:flat_direction}
%%%%%%%%%%%%%%%%%%%%%%%
%%%%%%%%%%%%%%%%%%%%%%%
%%%%%%%%%%%%%%%%%%%%%%%
Consider the $U(N)$ theory with $d$ matrices ($d=9$ in the case considered in this paper). 
First let us consider the simplest situation, where one of the D0-branes is separated by a distance 
$L$ from the other $(N-1)$ branes that are close to each other. By using the residual gauge symmetry 
and $SO(d)$ symmetry, we can take $X^d_{NN}\simeq L$ and have all other matrix components be much smaller than $L$.    
Then the potential energy coming from $N$-th row and column is approximately $\frac{1}{g^2}\sum_{i=1}^{d-1}\sum_{a=1}^{N-1} L^2 |X^i_{aN}|^2$ (here we are neglecting contributions from elements that do not scale with $L$). 
This contribution must be smaller than the total energy $E$ (which is kept fixed), and therefore possible values of $X^i_{aN}$ are suppressed as $L$ becomes large, as 
$\sum_{i=1}^{d-1}\sum_{a=1}^{N-1} |X^i_{aN}|^2\lesssim g^2 E/L^2$. Hence the phase space volume for $L>L_0$ is suppressed by at least
\begin{eqnarray}
\int_{L_0}^\infty \frac{L^{d-1}dL}{L^{2(d-1)(N-1)}}\sim \int_{L_0}^\infty \frac{dL}{L^{(d-1)(2N-3)}} \ed
\end{eqnarray}
The factor $L^{-2(d-1)(N-1)}$ comes from the integral with respect to $(d-1)(N-1)$ complex variables $X^i_{aN}$ $(a=1,\dots,N-1)$ and 
$L^{d-1}$ comes from $SO(d)$ rotational symmetry.  
As $L_0$ goes to infinity, the volume vanishes except for when $d=2$, $N=2$. In other words, this flat direction occupies a finite volume in phase space unless $d=2$, $N=2$.

When more eigenvalues are separated from the rest, more off-diagonal elements have to become small and hence such configurations have even smaller measures.


\begin{thebibliography}{99}

\bibitem{dankert:2009a}
C.~{Dankert}, R.~{Cleve}, J.~{Emerson}, and E.~{Livine}, ``Exact and
  approximate unitary 2-designs and their application to fidelity estimation,''
  Phys.Rev. {\bf A80} (2009), no.~1 012304,
	[\href{http://arxiv.org/abs/quant-ph/0606161}{quant-ph/0606161}].

\bibitem{harrow:2009a}
A.~W. {Harrow} and R.~A. {Low}, ``Random quantum circuits are approximate
  2-designs,''  Comm.Math.Phys. {\bf 291} (2009) 257--302,
	[\href{http://arxiv.org/abs/0802.1919}{arXiv:0802.1919} [quant-ph]].
 

\bibitem{arnaud:2008a}
L.~{Arnaud} and D.~{Braun}, ``Efficiency of producing random unitary
  matrices with quantum circuits,''  Phys.Rev. {\bf A78} (2008), no.~6
  062329, [\href{http://arxiv.org/abs/0807.0775}{arXiv:0807.0775} [quant-ph]].

\bibitem{brown:2010a}
W.~G. {Brown} and L.~{Viola}, ``Convergence rates for arbitrary statistical
  moments of random quantum circuits,''  Phys.Rev.Lett. {\bf 104} (2010),
  no.~25 250501, [\href{http://arxiv.org/abs/0910.0913}{arXiv:0910.0913} [quant-ph]].

\bibitem{diniz:2011a}
I.~T. Diniz and D.~Jonathan, ``Comment on the paper `{R}andom quantum
  circuits are approximate 2-designs',''  Comm.Math.Phys. {\bf 304}
  (2011) 281--293,  [\href{http://arxiv.org/abs/1006.4202}{arXiv:1006.4202} [quant-ph]].

\bibitem{Brown:2012gy} 
  W.~Brown and O.~Fawzi,
 ``Scrambling speed of random quantum circuits,''
  \href{http://arxiv.org/abs/1210.6644}{arXiv:1210.6644} [quant-ph].
  %%CITATION = ARXIV:1210.6644;%%

%\cite{Hayden:2007cs}
\bibitem{Hayden:2007cs} 
  P.~Hayden and J.~Preskill,
  ``Black holes as mirrors: Quantum information in random subsystems,''
  JHEP {\bf 0709}, 120 (2007)
  [\href{http://arxiv.org/abs/0708.4025}{arXiv:0708.4025} [hep-th]].
  %%CITATION = ARXIV:0708.4025;%%
  %170 citations counted in INSPIRE as of 28 Oct 2015
  
\bibitem{'tHooft:1990fr} 
  G.~'t Hooft,
  ``The black hole interpretation of string theory,''
  Nucl.\ Phys.\ B {\bf 335}, 138 (1990).
  %%CITATION = NUPHA,B335,138;%%

\bibitem{Kiem:1995iy} 
  Y.~Kiem, H.~L.~Verlinde and E.~P.~Verlinde,
  ``Black hole horizons and complementarity,''
  Phys.\ Rev.\ D {\bf 52}, 7053 (1995)
  [\href{http://arxiv.org/abs/hep-th/9502074}{hep-th/9502074}].
  %%CITATION = HEP-TH/9502074;%%

%\cite{Lashkari:2011yi}
\bibitem{Lashkari:2011yi} 
  N.~Lashkari, D.~Stanford, M.~Hastings, T.~Osborne and P.~Hayden,
  ``Towards the Fast Scrambling Conjecture,''
  JHEP {\bf 1304}, 022 (2013)
  doi:10.1007/JHEP04(2013)022
  [\href{http://arxiv.org/abs/1111.6580}{arXiv:1111.6580} [hep-th]].
  %%CITATION = doi:10.1007/JHEP04(2013)022;%%
  %44 citations counted in INSPIRE as of 24 Nov 2015

%\cite{Sekino:2008he}
\bibitem{Sekino:2008he} 
  Y.~Sekino and L.~Susskind,
  ``Fast Scramblers,''
  JHEP {\bf 0810}, 065 (2008)
  [\href{http://arxiv.org/abs/0808.2096}{arXiv:0808.2096} [hep-th]].
  %%CITATION = ARXIV:0808.2096;%%
  %159 citations counted in INSPIRE as of 26 Oct 2015
  
    %\cite{deWit:1988ig}
\bibitem{deWit:1988ig} 
  B.~de Wit, J.~Hoppe and H.~Nicolai,
 ``On quantum mechanics of supermembranes,"
  Nucl.\ Phys.\ B {\bf 305}, 545 (1988).
  %%CITATION = NUPHA,B305,545;%%
  %587 citations counted in INSPIRE as of 03 Aug 2013



  %\cite{Banks:1996vh}
\bibitem{Banks:1996vh}
T.~Banks, W.~Fischler, S.~H.~Shenker and L.~Susskind,
``M theory as a matrix model: A conjecture,''
Phys.\ Rev.\ D {\bf 55}, 5112 (1997).     
[arXiv:hep-th/9610043].
%%CITATION = HEP-TH 9610043;%%


%\cite{Itzhaki:1998dd}
\bibitem{Itzhaki:1998dd}
  N.~Itzhaki, J.~M.~Maldacena, J.~Sonnenschein and S.~Yankielowicz,
``Supergravity and the large N limit of theories with sixteen
  supercharges',
Phys.\ Rev.\ D {\bf 58}, 046004 (1998).
% \prd{58}{1998}{046004} [\hepth{9802042}].
  %%CITATION = HEP-TH 9802042;%%


%\cite{Shenker:2013pqa}
\bibitem{Shenker:2013pqa} 
  S.~H.~Shenker and D.~Stanford,
  ``Black holes and the butterfly effect,''
  JHEP {\bf 1403}, 067 (2014)
  [\href{http://arxiv.org/abs/1306.0622}{arXiv:1306.0622} [hep-th]].
  %%CITATION = ARXIV:1306.0622;%%
  %43 citations counted in INSPIRE as of 24 gen 2015

%\cite{Shenker:2013yza}
\bibitem{Shenker:2013yza} 
  S.~H.~Shenker and D.~Stanford,
  ``Multiple Shocks,''
  JHEP {\bf 1412}, 046 (2014)
  [\href{http://arxiv.org/abs/1312.3296}{arXiv:1312.3296} [hep-th]].
  %%CITATION = ARXIV:1312.3296;%%
  %39 citations counted in INSPIRE as of 07 Nov 2015

%\cite{Shenker:2014cwa}
\bibitem{Shenker:2014cwa} 
  S.~H.~Shenker and D.~Stanford,
  ``Stringy effects in scrambling,''
  JHEP {\bf 1505}, 132 (2015)
  [\href{http://arxiv.org/abs/1412.6087}{arXiv:1412.6087} [hep-th]].
  %%CITATION = ARXIV:1412.6087;%%
  %15 citations counted in INSPIRE as of 26 Oct 2015

%\cite{Roberts:2014isa}
\bibitem{Roberts:2014isa} 
  D.~A.~Roberts, D.~Stanford and L.~Susskind,
  ``Localized shocks,''
  JHEP {\bf 1503}, 051 (2015)
  [\href{http://arxiv.org/abs/1409.8180}{arXiv:1409.8180} [hep-th]].
  %%CITATION = ARXIV:1409.8180;%%
  %20 citations counted in INSPIRE as of 07 Nov 2015

%\cite{Kitaev_talk_2014}
  \bibitem{Kitaev_talk_2014}
  A. Kitaev, 
  ``Hidden Correlations in the Hawking Radiation and Thermal Noise," talk
given at Fundamental Physics Prize Symposium, Nov. 10, 2014.
Stanford SITP seminars, Nov. 11 and Dec. 18, 2014.

%\cite{Maldacena:2015waa}
\bibitem{Maldacena:2015waa} 
  J.~Maldacena, S.~H.~Shenker and D.~Stanford,
  ``A bound on chaos,''
  \href{http://arxiv.org/abs/1503.01409}{arXiv:1503.01409} [hep-th].
  %%CITATION = ARXIV:1503.01409;%%
  %22 citations counted in INSPIRE as of 29 Oct 2015

%\cite{Larkin1969}
\bibitem{Larkin1969} 
  A. I. Larkin and Y. N. Ovchinnikov, 
  ``Quasiclassical method in the theory of super-conductivity," JETP 28, 6 (1969):  1200-1205

%\cite{Almheiri:2013hfa}
\bibitem{Almheiri:2013hfa} 
  A.~Almheiri, D.~Marolf, J.~Polchinski, D.~Stanford and J.~Sully,
  ``An Apologia for Firewalls,''
  JHEP {\bf 1309}, 018 (2013)
  [\href{http://arxiv.org/abs/1304.6483}{arXiv:1304.6483} [hep-th]].
  %%CITATION = ARXIV:1304.6483;%%
  %148 citations counted in INSPIRE as of 09 Nov 2015

%\cite{Polchinski:1999yd}
\bibitem{Polchinski:1999yd}
  J.~Polchinski, L.~Susskind and N.~Toumbas,
  ``Negative energy, superluminosity and holography,''
  Phys.\ Rev.\ D {\bf 60} (1999) 084006
  [\href{http://arxiv.org/abs/hep-th/9903228}{hep-th/9903228}].
  %%CITATION = HEP-TH/9903228;%%
  %78 citations counted in INSPIRE as of 07 Nov 2015

%\cite{Kitaev_talk_2015}
  \bibitem{Kitaev_talk_2015}

``A simple model of quantum holography," 
talk given at KITP Program: Entanglement in Strongly-Correlated Quantum Matter, 
Apr 7 and May 27, 2015.

%\cite{Stanford_talk_2015}
  \bibitem{Stanford_talk_2015}
``A bound on chaos," 
talk given at SITP Workshop: Fundamental bounds on quantum dynamics: Chaos, Dissipation, Entanglement, and Complexity, Oct 17, 2015.

%\cite{Sachdev:1992fk}
\bibitem{Sachdev:1992fk} 
  S.~Sachdev and J.~w.~Ye,
  ``Gapless spin fluid ground state in a random, quantum Heisenberg magnet,''
  Phys.\ Rev.\ Lett.\  {\bf 70}, 3339 (1993)
  [\href{http://arxiv.org/abs/cond-mat/9212030}{cond-mat/9212030}].
  %%CITATION = COND-MAT/9212030;%%
  %17 citations counted in INSPIRE as of 07 Nov 2015

%\cite{Sachdev:2015efa}
\bibitem{Sachdev:2015efa} 
  S.~Sachdev,
  ``Bekenstein-Hawking Entropy and Strange Metals,''
  \href{http://arxiv.org/abs/1506.05111}{arXiv:1506.05111} [hep-th].
  %%CITATION = ARXIV:1506.05111;%%
  %5 citations counted in INSPIRE as of 07 Nov 2015


%\cite{Matinyan:1981dj}
\bibitem{Matinyan:1981dj} 
  S.~G.~Matinyan, G.~K.~Savvidy and N.~G.~Ter-Arutunian Savvidy,
  ``Classical Yang-mills Mechanics. Nonlinear Color Oscillations,''
  Sov.\ Phys.\ JETP {\bf 53}, 421 (1981)
  [Zh.\ Eksp.\ Teor.\ Fiz.\  {\bf 80}, 830 (1981)].
  %%CITATION = SPHJA,53,421;%%
  %124 citations counted in INSPIRE as of 07 Nov 2015


  %\cite{Savvidy:1982wx}
\bibitem{Savvidy:1982wx}
  G.~K.~Savvidy,
  ``Yang-mills Classical Mechanics As A Kolmogorov K System,''
  Phys.\ Lett.\ B {\bf 130} (1983) 303.
  %%CITATION = PHLTA,B130,303;%%
  %62 citations counted in INSPIRE as of 07 Nov 2015


  %\cite{Savvidy:1982jk}
\bibitem{Savvidy:1982jk} 
  G.~K.~Savvidy,
  ``Classical and Quantum Mechanics of Nonabelian Gauge Fields,''
  Nucl.\ Phys.\ B {\bf 246}, 302 (1984).
  %%CITATION = NUPHA,B246,302;%%
  %81 citations counted in INSPIRE as of 07 Nov 2015


%\cite{Aref'eva:1997es}
\bibitem{Aref'eva:1997es} 
  I.~Y.~Aref'eva, P.~B.~Medvedev, O.~A.~Rytchkov and I.~V.~Volovich,
  ``Chaos in M(atrix) theory,''
  Chaos Solitons Fractals {\bf 10}, 213 (1999)
  %doi:10.1016/S0960-0779(98)00159-3
  [\href{http://arxiv.org/abs/hep-th/9710032}{hep-th/9710032}].
  %%CITATION = doi:10.1016/S0960-0779(98)00159-3;%%
  %22 citations counted in INSPIRE as of 09 Jan 2016




%\cite{Aref'eva:1998mk}
\bibitem{Aref'eva:1998mk} 
  I.~Y.~Aref'eva, A.~S.~Koshelev and P.~B.~Medvedev,
  ``Chaos order transition in Matrix theory,''
  Mod.\ Phys.\ Lett.\ A {\bf 13}, 2481 (1998)
  %doi:10.1142/S0217732398002643
  [\href{http://arxiv.org/abs/hep-th/9804021}{hep-th/9804021}].
  %%CITATION = doi:10.1142/S0217732398002643;%%
  %10 citations counted in INSPIRE as of 09 Jan 2016

%\cite{Asplund:2011qj}
\bibitem{Asplund:2011qj} 
  C.~Asplund, D.~Berenstein and D.~Trancanelli,
  ``Evidence for fast thermalization in the plane-wave matrix model,''
  Phys.\ Rev.\ Lett.\  {\bf 107}, 171602 (2011)
  [\href{http://arxiv.org/abs/1104.5469}{arXiv:1104.5469} [hep-th]].
  %%CITATION = ARXIV:1104.5469;%%
  %38 citations counted in INSPIRE as of 07 Nov 2015

%\cite{Asplund:2012tg}
\bibitem{Asplund:2012tg} 
  C.~T.~Asplund, D.~Berenstein and E.~Dzienkowski,
  ``Large N classical dynamics of holographic matrix models,''
  Phys.\ Rev.\ D {\bf 87}, no. 8, 084044 (2013)
  [\href{http://arxiv.org/abs/1211.3425}{arXiv:1211.3425} [hep-th]].
  %%CITATION = ARXIV:1211.3425;%%
  %24 citations counted in INSPIRE as of 07 Nov 2015

%\cite{Asano:2015eha}
\bibitem{Asano:2015eha} 
  Y.~Asano, D.~Kawai and K.~Yoshida,
  ``Chaos in the BMN matrix model,''
  JHEP {\bf 1506}, 191 (2015)
  [\href{http://arxiv.org/abs/1503.04594}{arXiv:1503.04594} [hep-th]].
  %%CITATION = ARXIV:1503.04594;%%
  %2 citations counted in INSPIRE as of 07 Nov 2015

%\cite{Aoki:2015uha}
\bibitem{Aoki:2015uha} 
  S.~Aoki, M.~Hanada and N.~Iizuka,
  %``Quantum Black Hole Formation in the BFSS Matrix Model,''
  JHEP {\bf 1507}, 029 (2015)
  [\href{http://arxiv.org/abs/1503.05562}{arXiv:1503.05562} [hep-th]].
  %%CITATION = ARXIV:1503.05562;%%


  %\cite{Anagnostopoulos:2007fw}
\bibitem{Anagnostopoulos:2007fw} 
  K.~N.~Anagnostopoulos, M.~Hanada, J.~Nishimura and S.~Takeuchi,
 ``Monte Carlo studies of supersymmetric matrix quantum mechanics with sixteen supercharges at finite temperature,''
  Phys.\ Rev.\ Lett.\  {\bf 100}, 021601 (2008). 
%  [\href{http://arxiv.org/abs/0707.4454}{arXiv:0707.4454} [hep-th]].
  %%CITATION = ARXIV:0707.4454;%%

%\cite{Catterall:2008yz}
%\bibitem{Catterall:2008yz}
  S.~Catterall and T.~Wiseman,
``Black hole thermodynamics from simulations 
 of lattice Yang-Mills theory,''
  Phys.\ Rev.\  D {\bf 78}, 041502 (2008). 
%  [\href{http://arxiv.org/abs/0803.4273}{arXiv:0803.4273} [hep-th]].
  %%CITATION = PHRVA,D78,041502;%%

  %\cite{Hanada:2008ez}
%\bibitem{Hanada:2008ez} 
  M.~Hanada, Y.~Hyakutake, J.~Nishimura and S.~Takeuchi,
  ``Higher derivative corrections to black hole thermodynamics from supersymmetric matrix quantum mechanics,''
  Phys.\ Rev.\ Lett.\  {\bf 102}, 191602 (2009). 
%  [\href{http://arxiv.org/abs/0811.3102}{arXiv:0811.3102} [hep-th]].
  %%CITATION = ARXIV:0811.3102;%%  

%\cite{Kadoh:2015mka}
%\bibitem{Kadoh:2015mka} 
  D.~Kadoh and S.~Kamata,
  ``Gauge/gravity duality and lattice simulations of one dimensional SYM with sixteen supercharges,''
  \href{http://arxiv.org/abs/1503.08499}{arXiv:1503.08499} [hep-lat].
  %%CITATION = ARXIV:1503.08499;%%
  %4 citations counted in INSPIRE as of 09 Nov 2015


%\cite{Filev:2015hia}
%\bibitem{Filev:2015hia} 
  V.~G.~Filev and D.~O'Connor,
  ``The BFSS model on the lattice,''
  \href{http://arxiv.org/abs/1506.01366}{arXiv:1506.01366} [hep-th].
  %%CITATION = ARXIV:1506.01366;%%
  %1 citations counted in INSPIRE as of 09 Nov 2015


  %\cite{Witten:1995im}
\bibitem{Witten:1995im} 
  E.~Witten,
  ``Bound states of strings and p-branes,''
  Nucl.\ Phys.\ B {\bf 460}, 335 (1996)
  [\href{http://arxiv.org/abs/hep-th/9510135}{hep-th/9510135}].
  %%CITATION = HEP-TH/9510135;%%
  %1242 citations counted in INSPIRE as of 04 Dec 2014

%\cite{Kawahara:2007ib}
\bibitem{Kawahara:2007ib}
  N.~Kawahara, J.~Nishimura and S.~Takeuchi,
  ``High temperature expansion in supersymmetric matrix quantum mechanics,''
  JHEP {\bf 0712} (2007) 103
  [\href{http://arxiv.org/abs/0710.2188}{arXiv:0710.2188} [hep-th]].
  %%CITATION = ARXIV:0710.2188;%%
  %22 citations counted in INSPIRE as of 31 May 2015

%\cite{oseledets1968multiplicative}
\bibitem{oseledets1968multiplicative}
Valery~Iustinovich Oseledets.
\newblock A multiplicative ergodic theorem. characteristic ljapunov, exponents
  of dynamical systems.
\newblock {\em Trudy Moskovskogo Matematicheskogo Obshchestva}, 19:179--210,
  1968.

%\cite{Xu20031}
\bibitem{Xu20031}
Hongguo Xu,
\newblock An SVD-Like Matrix Decomposition and its Applications,
\newblock {\em Linear Algebra and its Applications}, 368:1 -- 24, 2003.

%\cite{goldhirsch1987stability}
\bibitem{goldhirsch1987stability}
Isaac Goldhirsch, Pierre-Louis Sulem, and Steven~A Orszag.
\newblock Stability and lyapunov stability of dynamical systems: A differential
  approach and a numerical method.
\newblock {\em Physica D: Nonlinear Phenomena}, 27(3):311--337, 1987.

%\cite{sprott2003chaos}
\bibitem{sprott2003chaos}
  J.~C.~Sprott,
  \newblock {\em Chaos and Time-Series Analysis}, 
  \newblock Oxford University Press, 2003.

%\cite{PhysRevA.66.052302}
\bibitem{PhysRevA.66.052302}
Robert Alicki.
\newblock Information-theoretical meaning of quantum-dynamical entropy.
\newblock {\em Phys. Rev. A}, 66:052302, Nov 2002.

%\cite{alicki1998quantum}
\bibitem{alicki1998quantum}
Robert Alicki.
\newblock Quantum geometry of noncommutative bernoulli shifts.
\newblock {\em Banach Center Publications}, 43(1):25--29, 1998.

%\cite{mendes1995entropy}
\bibitem{mendes1995entropy}
R~Vilela Mendes.
\newblock Entropy and quantum characteristic exponents. steps towards a quantum
  pesin theory.
\newblock In {\em Chaos—The Interplay between Stochastic and Deterministic
  Behaviour}, pages 273--282. Springer, 1995.

%\cite{man2000lyapunov}
\bibitem{man2000lyapunov}
VI~Man’ko and R~Vilela Mendes.
\newblock Lyapunov exponent in quantum mechanics. a phase-space approach.
\newblock {\em Physica D: Nonlinear Phenomena}, 145(3):330--348, 2000.

%\cite{man2002quantum}
\bibitem{man2002quantum}
VI~Man'ko and R~Vilela Mendes.
\newblock Quantum sensitive dependence.
\newblock {\em Physics Letters A}, 300(4):353--360, 2002.


%\cite{Kunihiro01032009}
\bibitem{Kunihiro01032009}
Teiji Kunihiro, Berndt Müller, Akira Ohnishi, and Andreas Schäfer.
\newblock Towards a theory of entropy production in the little and big bang.
\newblock {\em Progress of Theoretical Physics}, 121(3):555--575, 2009.

%\cite{Asplund:2015osa}
\bibitem{Asplund:2015osa} 
  C.~T.~Asplund and D.~Berenstein,
  ``Entanglement entropy converges to classical entropy around periodic orbits,''
  \href{http://arxiv.org/abs/1503.04857}{arXiv:1503.04857} [hep-th].
  %%CITATION = ARXIV:1503.04857;%%
  %2 citations counted in INSPIRE as of 07 Nov 2015


%\cite{Zurek:1994wd}
\bibitem{Zurek:1994wd} 
  W.~H.~Zurek and J.~P.~Paz,
  ``Decoherence, chaos, and the second law,''
  Phys.\ Rev.\ Lett.\  {\bf 72}, 2508 (1994)
  [\href{http://arxiv.org/abs/gr-qc/9402006}{gr-qc/9402006}].
  %%CITATION = GR-QC/9402006;%%
  %77 citations counted in INSPIRE as of 07 Nov 2015

%\cite{0264-9381-25-19-195005}
\bibitem{0264-9381-25-19-195005}
Kostyantyn Ropotenko.
\newblock Kolmogorov–sinai entropy and black holes.
\newblock {\em Classical and Quantum Gravity}, 25(19):195005, 2008.

%\cite{Balasubramanian:2011ur}
\bibitem{Balasubramanian:2011ur} 
  V.~Balasubramanian {\it et al.},
  ``Holographic Thermalization,''
  Phys.\ Rev.\ D {\bf 84}, 026010 (2011)
  [\href{http://arxiv.org/abs/1103.2683}{arXiv:1103.2683} [hep-th]].
  %%CITATION = ARXIV:1103.2683;%%
  %149 citations counted in INSPIRE as of 07 Nov 2015

%\cite{Li2013}
\bibitem{Li2013}
Yong-Zhuang Li, Shao-Feng Wu, Yong-Qiang Wang, and Guo-Hong Yang.
\newblock Linear growth of entanglement entropy in holographic thermalization
  captured by horizon interiors and mutual information.
\newblock {\em Journal of High Energy Physics}, 2013(9).

\bibitem{stanfordUnpub}
  D. Stanford, unpublished.

%\cite{Liu:2013iza}
\bibitem{Liu:2013iza} 
  H.~Liu and S.~J.~Suh,
  ``Entanglement Tsunami: Universal Scaling in Holographic Thermalization,''
  Phys.\ Rev.\ Lett.\  {\bf 112}, 011601 (2014)
  [\href{http://arxiv.org/abs/1305.7244}{arXiv:1305.7244} [hep-th]].
  %%CITATION = ARXIV:1305.7244;%%
  %55 citations counted in INSPIRE as of 26 Oct 2015

%\cite{eichhorn2001transformation}
\bibitem{eichhorn2001transformation}
Ralf Eichhorn, Stefan~J Linz, and Peter H{\"a}nggi.
\newblock Transformation invariance of lyapunov exponents.
\newblock {\em Chaos, Solitons \& Fractals}, 12(8):1377--1383, 2001.

\end{thebibliography}
\end{document}